\documentclass[%
reprint,
onecolumn,
longbibliography,
 amsmath,amssymb,
 aps,
 showkeys,
]{revtex4-2}

\usepackage{siunitx}
\usepackage{graphicx}
\usepackage{dcolumn}
\usepackage{bm}
\usepackage[caption=false]{subfig}
\usepackage{tikz}

\begin{document}

\preprint{APS/123-QED}

\title{A numerical study of a pair of spheres in an oscillating box filled with viscous fluid}

\author{T.J.J.M. van Overveld}
\affiliation{Fluids and Flows group and J.M. Burgers Center for Fluid Dynamics, Department of Applied Physics, Eindhoven University of Technology, P.O. Box 513, 5600 MB Eindhoven, The Netherlands}
\author{M.T. Shajahan}
\affiliation{Laboratory for Aero and Hydrodynamics, Delft University of Technology, 2628 CD Delft, The Netherlands}
\author{W.-P. Breugem}
\affiliation{Laboratory for Aero and Hydrodynamics, Delft University of Technology, 2628 CD Delft, The Netherlands}
\author{H.J.H. Clercx}
\affiliation{Fluids and Flows group and J.M. Burgers Center for Fluid Dynamics, Department of Applied Physics, Eindhoven University of Technology, P.O. Box 513, 5600 MB Eindhoven, The Netherlands }
\author{M. Duran-Matute}%
\email{m.duran.matute@tue.nl}
\affiliation{Fluids and Flows group and J.M. Burgers Center for Fluid Dynamics, Department of Applied Physics, Eindhoven University of Technology, P.O. Box 513, 5600 MB Eindhoven, The Netherlands }

\date{\today}

\begin{abstract}
    When two spherical particles submerged in a viscous fluid are subjected to an oscillatory flow, they align themselves perpendicular to the direction of the flow leaving a small gap between them. The formation of this compact structure is attributed to a non-zero residual flow known as \emph{steady streaming}.
    We have performed direct numerical simulations of a fully-resolved, oscillating flow in which the pair of particles is modeled using an immersed boundary method. Our simulations show that the particles oscillate both parallel and perpendicular to the oscillating flow in elongated \emph{figure 8} trajectories. In absence of bottom friction, the mean gap between the particles depends only on the normalized Stokes boundary layer thickness $\delta^*$, and on the normalized, streamwise excursion length of the particles relative to the fluid $A_r^*$ (equivalent to the Keulegan-Carpenter number). For $A_r^*\lesssim 1$, viscous effects dominate and the mean particle separation only depends on $\delta^*$. For larger $A_r^*$-values, advection becomes important and the gap widens. Overall, the normalized mean gap between the particles scales as $L^*\approx3.0{\delta^*}^{1.5}+0.03{A_r^*}^3$, which also agrees well with previous experimental results. The two regimes are also observed in the magnitude of the oscillations of the gap perpendicular to the flow, which increases in the viscous regime and decreases in the advective regime. When bottom friction is considered, particle rotation increases and the gap widens. Our results stress the importance of simulating the particle motion with all its degrees of freedom to accurately model the system and reproduce experimental results. The new insights of the particle pairs provide an important step towards understanding denser and more complex systems.
\end{abstract}

\maketitle

\section{\label{sec:Introduction}Introduction}
    The formation of patterns in granular systems is a widespread phenomenon that is commonly encountered in maritime environments, such as bed ripples under waves \cite{Blondeaux1990,Bagnold1946}. Understanding the evolution of such structures can be crucial to protect sensitive coastal areas by predicting e.g. the bed erosion \cite{Vittori2020,Kidanemariam2014}. Additionally, the emergence of patterns can be exploited for multiple industrial applications \cite{Jaeger1996,Aranson2006}, such as the separation of mixtures with multiple granular components \cite{Sanchez2004}.
    
    One particular example of pattern formation occurs when a few spherical particles are placed in a fluid-filled cell undergoing horizontal, sinusoidal vibration. The spheres form one-particle-thick chains that align perpendicularly to the oscillation direction with equal spacing between the chains \cite{Wunenburger2002,Klotsa2009}. This phenomenon is driven by a non-zero residual flow that remains when averaging over a full oscillation period. This residual, known as ‘steady streaming’, is a result of the non-linear interaction between the flow and the particles \cite{Riley1966,Wang1965,Andres1953,Andres1953a}.
    
    The steady streaming flow around a set of (unequally sized) particles can generate net propulsion \cite{Klotsa2015,Dombrowski2020}. This is interesting from a biological point of view since there is resemblance with swimming organisms at moderate Reynolds numbers \cite{Leoni2009}. This knowledge can be further exploited in the development of efficient, passive robots for drug delivery in the human body \cite{Feng2014}. 
    
    A fundamental approach to understand the particle-fluid interactions that drive these systems is to consider a pair of identical spheres. In a symmetrically oscillating flow, they form a stable side-by-side configuration with a gap between them. Such a pair can be seen as the smallest possible chain and the building block for longer ones \cite{Klotsa2009}.
    \citeauthor{Klotsa2007}~\cite{Klotsa2007} studied this phenomenon experimentally by placing a couple of spheres in a vibrating box at high frequencies and small amplitudes. 
    They linked the existence of an equilibrium gap to the balance between long-range attractive and short-range repulsive forces between the particles. This implies that there is a stable point where the net spanwise force on the spheres is zero. However, due to the fact that this force is the result of non-linear interactions, the position of a second particle cannot be derived from the average flow around a single particle.
    
    Furthermore, they proposed two distinct scalings of the gap as a function of the flow conditions, with a relatively abrupt transition between the scalings. The location of the transition in the parameter space was found to depend on both the oscillatory boundary layer thickness and the relative excursion length of the particle with respect to the fluid. The regimes were distinguished by qualitative changes in the steady streaming flow, obtained from numerical simulations. Quantitatively, these simulations overestimated the gap by approximately a factor of two. 
    In the numerical simulations, the fluid was at rest, while the particles moved along a predefined, sinusoidal path in one coordinate direction. This method allows full control of particle excursion lengths and is useful for comparison with experimental data. However, it does not account for deviations from the harmonic motion in the driving direction, which may occur as a result of interactions with boundaries or the steady streaming flow.
    Moreover, this driving method is subtly different from applying a sinusoidal force on the particles and let their motion emerge. The latter driving method corresponds to the system in the reference frame moving along with the box.
    
    In related numerical studies \cite{Jalal2016,Fabre2017}, the average force between a pair of fixed spheres was used to estimate the equilibrium spacing between the pair. However, this was done for an infinite domain (i.e. in absence of solid boundaries such as an impermeable no-slip bottom) and for limiting cases where the amplitude of the oscillating flow goes to zero. These results correspond thus only to a limited part of the parameter space.
    
    In the aforementioned numerical studies on particle pairs (see Refs. \cite{Klotsa2007,Klotsa2009,Jalal2016,Fabre2017}), the motion of the particles is restricted in one or more directions. As of present, there are no numerical studies of this oscillating box system, in which the motion of the particles is obtained directly as a result of the particle-fluid interactions. 
    
    In this study, we performed direct numerical simulations using an immersed boundary method (IBM) to simulate the motion of the particles. We do thus not impose any restrictions to the particle motion, but let their dynamics evolve only from interaction with the fluid.
    We focus on the fundamental aspects regarding particle pairs, such as their dynamics and the gap size between them as a function of the problem parameters. Due to computational advances, our simulations allow us to considerably extend the parameter space investigated by \citeauthor{Klotsa2007}~\cite{Klotsa2007}, and explore richer particle dynamics than previously observed. In particular, we observe and describe, for the first time, an oscillation of the particles in the cross-flow direction. Using the simulation results, we further propose a new scaling of the mean gap as a function of the parameters of the problem. The new scaling indicates the existence of two regimes, which we explain using the steady streaming flow around the pair. We further show that the gap is governed by only two dimensionless parameters, by combining the results with dimensional analysis and non-dimensionalization of the governing equations.
    
    This paper is organized as follows. We start with a definition of the physical systems and identify the relevant parameters, equations and dimensionless quantities in Sec.~\ref{sec:System}. In Sec.~\ref{sec:Method}, the numerical approach based on the immersed boundary method (IBM) by \citeauthor{Breugem2012}~\cite{Breugem2012} is described. The results are presented in Sec.~\ref{sec:Results} and discussed in Sec.~\ref{sec:Discussion}. The effects of particle rotation and inertia are addressed separately. We end with conclusions in Sec.~\ref{sec:Conclusions}.

\section{\label{sec:System}Description of the physical system}
\subsection{Overview of the system}
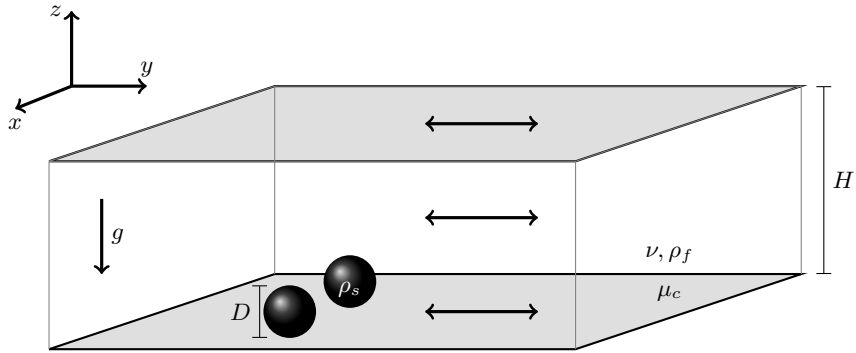
\begin{figure}
        \centering
        \begin{tikzpicture}
            \def\w{7};   
            \def\xl{3};
            \def\yl{1};
            \def\h{2.5}; 
            
            \filldraw[fill=lightgray!50, thick] (0,0) -- (\w,0) -- (\w+\xl,\yl) -- (\xl,\yl) -- (0,0);
            
            \filldraw[fill=lightgray!50, thick] (0,\h) -- (\w,\h) -- (\w+\xl,\h+\yl) -- (\xl,\h+\yl) -- (0,\h);
            
            \shade[ball color = black] (3.2,0.5) circle (0.35); 
            \draw[|-|] (2.8,0.15) -- node[anchor=east] {$D$} (2.8,0.85);
            \shade[ball color = black] (4,0.9) circle (0.35); 
            \node[color=white] at (4,0.8) {$\rho_s$};

            \draw[gray] (0,\h) -- (\w,\h) -- (\w+\xl,\yl+\h) -- (\xl,\yl+\h) -- (0,\h);
            \draw[gray] (0,0) -- (0,\h);
            \draw[gray] (\w,0) -- (\w,\h);
            \draw[gray] (\w+\xl,\yl) -- (\w+\xl,\yl+\h);
            \draw[gray] (\xl,\yl) -- (\xl,\yl+\h);
            
            \draw[|-|] (\w+\xl+0.3,\yl) -- node[anchor=west] {$H$} (\w+\xl+0.3,\h+\yl);
            
            \coordinate (O) at (0.3,3.5);
            
            \draw[->,very thick] (O) -- ++(-0.75,-0.3) node[anchor=north] {$x$};
            \draw[->,very thick] (O) -- ++(1,0) node[anchor=south] {$y$};
            \draw[->,very thick] (O) -- ++(0,1) node[anchor=east] {$z$};
            
            \draw[->,very thick] (0.7,2) -- node[anchor=west] {$g$} (0.7,1) ;
            
            \draw[<->,very thick] (5,0.5) -- (6.5,0.5) ;
            \draw[<->,very thick] (5,1.75) -- (6.5,1.75) ;
            \draw[<->,very thick] (5,3.0) -- (6.5,3.0) ;
            
            \node at (8.25,1.25) {$\nu,\rho_f$};
            \node at (8.25,0.75) {$\mu_c$};
            
            \end{tikzpicture}
        \caption{A schematic overview of the system considered in this study, along with the relevant dimensionful quantities of Table~\ref{tab:quantities}. The system is shown in the lab frame, in which the bottom, top, and fluid oscillate in the $y$-direction, as indicated by the double headed arrows.}
        \label{fig:systemsketch}
    \end{figure}

    The physical system consists of a pair of submerged spherical particles with diameter $D$ on top of an infinitely large, horizontal bottom plate. A second plate is placed parallel to the bottom at a height $H$. The fluid between the plates has density $\rho_f$ and kinematic viscosity $\nu$. The two spherical particles have a density $\rho_s$ such that $\rho_s>\rho_f$. Due to gravity, the particles are always in contact with the bottom plate. The Coulomb friction coefficient between the particles and bottom is $\mu_c$. 
    An external pressure gradient imposed on the fluid drives an oscillating flow with excursion length $A$ and angular frequency $\omega$. The corresponding oscillation velocity amplitude of the bulk flow is $A\omega$. The top and bottom plates move with the same amplitude and frequency as the fluid, such that the system represents an oscillating box.
    A Cartesian coordinate system is defined with the $z$\nobreakdash-coordinate in the vertical direction, perpendicular to the bottom. The $y$\nobreakdash-axis is parallel to the oscillation direction, and the $x$\nobreakdash-axis is in the other horizontal direction. The dimensionful quantities defining the problem and their units are summarized in Table~\ref{tab:quantities}, and a schematic overview is given in Fig.~\ref{fig:systemsketch}.

    \begin{table}
        \caption{An overview of the dimensionful quantities defining the system with their symbols and units.}\label{tab:quantities}
        \begin{ruledtabular}
        \begin{tabular}{ccc}
        Symbol & Description & SI units\\ \hline
        $D$ & Particle diameter & \SI{}{m}\\
        $H$ & Plate separation & \SI{}{m}\\
        $A$ & Oscillation amplitude & \SI{}{m}\\
        $\omega$ & Oscillation frequency & \SI{}{(rad)/s}\\
        $\rho_f$ & Fluid density & \SI{}{kg/m^3}\\
        $\rho_s$ & Particle density & \SI{}{kg/m^3}\\
        $\nu$ & Kinematic viscosity & \SI{}{m^2/s}\\
        $g$ & Gravitational acceleration & \SI{}{m/s^2}\\
        $\mu_c$ & Coulomb friction coefficient & \SI{}{-} \\
        \end{tabular}
        \end{ruledtabular}
    \end{table}
    
    Based on the aforementioned quantities, dimensional analysis yields six dimensionless numbers that characterize the problem. We choose the closed set consisting of the amplitude (or excursion length) of the oscillating flow normalized by the particle diameter
    \begin{equation}
        A^*=\frac{A}{D},
    \end{equation}
    the distance between the plates normalized by the particle diameter
    \begin{equation}
        H^*=\frac{H}{D},
    \end{equation}
    the ratio between the Stokes boundary layer thickness and particle size
    \begin{equation}
        \delta^*=\frac{\delta}{D} = \frac{\sqrt{2\nu/\omega}}{D},
    \end{equation}
    the particle-fluid density ratio
    \begin{equation}
        s=\frac{\rho_s}{\rho_f},
    \end{equation}
    the ratio between the oscillatory and gravitational acceleration
    \begin{equation}
        \Gamma = \frac{A\omega^2}{g},
    \end{equation}
    and the Coulomb friction coefficient $\mu_c$. The viscous boundary layers with thickness $\delta$ are present near the no\nobreakdash-slip boundaries in the domain. In this specific system, they are only present around the spheres, since the top and bottom are oscillating in unison with the fluid. We restrict ourselves to the case $H^*\gg 1$, so that the upper boundary has no effect on the particles.
    
    In literature, different sets of dimensionless parameters are used to describe physically similar systems (see, e.g., Ref.~\cite{Klotsa2007}). With our choice of dimensionless numbers, other dimensionless parameters can be constructed as well, such as the Reynolds number of the oscillatory boundary layer
    \begin{equation}\label{reynoldsdelta}
        \mathrm{Re}_\delta = \frac{A\omega \delta}{\nu} =\frac{2A}{\delta}= \frac{2A^*}{\delta^*}.
    \end{equation}
    This particular quantity is relevant to the problem because it is a measure for the relative importance of inertial effects versus viscous drag on the scale of the Stokes boundary layer.

    The main output parameters of the system are related to the motion and relative position of the particles. The main motion of the particles is along the $y$\nobreakdash-axis, with excursion length $A_p$ in the lab reference frame or $A_r$ in the frame of the oscillating fluid. The pair lies predominantly in the $x$\nobreakdash-direction, with a mean gap size $L$. From here on the normalized variants of these quantities will be used, with $A_p^*=A_p/D$, $A_r^*=A_r/D$ and $L^*=L/D$.
    
\subsection{Governing equations}\label{sectionequations}
    \subsubsection{Fluid motion}
    The fluid is incompressible and Newtonian, so that the flow is governed by the continuity equation
    \begin{equation}
        \boldsymbol{\nabla}\cdot\boldsymbol{u} = 0,
    \end{equation}
    and the Navier-Stokes equation
    \begin{equation}\label{momentum}
        \frac{\partial \boldsymbol{u}}{\partial t} + \left(\boldsymbol{u}\cdot\boldsymbol{\nabla}\right)\boldsymbol{u} = -\frac{1}{\rho_f}\boldsymbol{\nabla} p + \nu\nabla^2\boldsymbol{u} + A\omega^2\cos\left(\omega t\right)\hat{\boldsymbol{y}},
    \end{equation}
    where $\boldsymbol{u}$ is the fluid velocity, $p$ the pressure and $t$ the time. The oscillating pressure gradient that drives the flow is written here explicitly as a body force in the $y$\nobreakdash-direction. The effect of gravity on the fluid is not considered since hydrostatic effects play no role and the system has no free surface. 
    At the bottom and top plate, the no\nobreakdash-slip/no\nobreakdash-penetration boundary condition
    \begin{equation}
        \left.\boldsymbol{u}\right|_{z=0,H}=A\omega\sin(\omega t)\hat{\boldsymbol{y}}
    \end{equation}
    holds, such that the fluid oscillates in unison with the boundaries.
    
    We introduce the typical velocity scale $A\omega$, the time scale $2\pi/\omega$, and the length scale $D$ to define the dimensionless variables, denoted with tildes, as follows:
    \begin{equation}\label{dimlesvars}
        \boldsymbol{u}=A\omega\tilde{\boldsymbol{u}}, \quad 
        t=\frac{2\pi}{\omega}\tilde{t}, \quad
        \left(x,y,z\right) = D\left(\tilde{x},\tilde{y},\tilde{z}\right), \quad
        \boldsymbol{\nabla} = \frac{1}{D}\tilde{\boldsymbol{\nabla}}, \quad
        p=\rho_f AD\omega^2\tilde{p}.
    \end{equation}
    Here the pressure $p$ is scaled such that the pressure gradient term in the dimensionless equation is of the same order as the local acceleration and forcing terms. 
    Reinserting Eq.~\eqref{dimlesvars} into Eq.~\eqref{momentum} and immediately dropping the tildes yields
    \begin{equation}\label{momentum_dimless}
        \frac{1}{2\pi}\frac{\partial \boldsymbol{u}}{\partial t} + A^*\left(\boldsymbol{u}\cdot\boldsymbol{\nabla}\right)\boldsymbol{u} = -\boldsymbol{\nabla} p + \frac{1}{2}\delta^{*2}\nabla^2\boldsymbol{u} + \cos\left(2\pi t\right)\hat{\boldsymbol{y}},
    \end{equation}
    which is equivalent to the formulation used for analysis of steady streaming flows by \citeauthor{Riley1966}~\cite{Riley1966}.
    In this non-dimensional equation, the main oscillating flow is governed by the balance between the local acceleration (first term on the left-hand side) and the forcing (last term on the right-hand side). The term $\boldsymbol{\nabla}p$ represents the dynamic pressure, which is non-zero when particles are present in the system.
    In absence of particles, the solution for the flow in the oscillating box is $\boldsymbol{u}=\sin(2\pi t)\hat{\boldsymbol{y}}$. This flow is symmetric and vanishes when it is averaged over a full oscillation period. Any additional flow structures, such as steady streaming flows, emerge when particles are added to the system. The evolution of such non-linear flows is governed by the normalized excursion length of the flow $A^*$ and the normalized viscous length scale $\delta^*$.
    
    Alternatively, we could choose $\delta$ as typical length scale in the non-dimensionalization, since variations in flow fields that are dominated by viscosity are typically expected on the scale of the oscillatory boundary layer thickness. This approach was used previously in the context of particle chains by \citeauthor{Mazzuoli2016}~\cite{Mazzuoli2016}, and yields
    \begin{equation}\label{momentum_dimless_alt}
        \frac{1}{2\pi}\frac{\partial \boldsymbol{u}}{\partial t} + \frac{A}{\delta}\left(\boldsymbol{u}\cdot\boldsymbol{\nabla}\right)\boldsymbol{u} = -\boldsymbol{\nabla} p + \frac{1}{2}\nabla^2\boldsymbol{u} + \cos\left(2\pi t\right)\hat{\boldsymbol{y}}.
    \end{equation}
    From this equation, we identify that the last two terms are of the same order of magnitude, which means that the viscous forces and the driving force should balance. In such case, the non-linearity in the flow is governed by the ratio $A/\delta$, which is the Reynolds number of the Stokes boundary layer [see Eq.~\eqref{reynoldsdelta}]. When the excursion length of the flow is large with respect to the boundary layer, i.e. $A/\delta\gg1$, the advection of structures in the flow field dominates. Because the advection term is the only non-linear term in Eq.~\eqref{momentum_dimless_alt}, we expect the importance of the steady streaming flow to scale with the same ratio $A/\delta$.
    In the remainder of this study, we follow the former approach using $D$ and Eq.~\eqref{momentum_dimless}, since it allows for a more straightforward comparison with fundamental analysis by e.g. \citeauthor{Riley1966}~\cite{Riley1966}.
    
    \subsubsection{Particle motion}
    Translation of the spherical particles in the lab frame is described by Newton's second law as
    \begin{equation}\label{particle_trans}
        \rho_s V_s \frac{d\boldsymbol{u}_s}{dt} = \oint\boldsymbol{\tau}\cdot\hat{\boldsymbol{n}}dS - \left(\rho_s-\rho_f\right)V_sg\hat{\boldsymbol{z}} + \boldsymbol{F}_c + \rho_f V_s A \omega^2\cos(\omega t)\hat{\mathbf{y}},
    \end{equation}
    where $V_s=\pi D^3/6$ is the volume of the particle and $\boldsymbol{F}_c$ represents collision forces due to interactions with other particles or boundaries. The interaction with the fluid is accounted for by integrating the stress tensor $\boldsymbol{\tau}$ over the surface of the sphere, here written as $S$. The outward vector normal to the surface of the spherical particle is $\hat{\boldsymbol{n}}$. Both the hydrodynamic drag and lift forces ($\boldsymbol{F}_L$) are incorporated in this integral. 
    The last term on the right-hand side is the body force due to the pressure gradient in the undisturbed flow, given by $\rho_f V_s (d\boldsymbol{u}/dt)$. Hence, it is similar to the last term of Eq.~\eqref{momentum}.
    
    In all cases considered in this study, the particles are continuously touching the horizontal bottom. This means that the normal force $\boldsymbol{F}_c\cdot\hat{\boldsymbol{z}}$ is such that the net vertical force on the particles is zero. In other words, the $z$\nobreakdash-component of the collision force counteracts the gravity, buoyancy and lift forces, such that
    \begin{equation}
         \boldsymbol{F}_c\cdot\hat{\boldsymbol{z}} = \left(\rho_s-\rho_f\right)V_s g-\boldsymbol{F}_L\cdot\hat{\boldsymbol{z}}.
    \end{equation}

    Besides the bottom, the particles do not touch other solid surfaces, i.e. particles do not collide with each other. Hence, the horizontal components of $\boldsymbol{F}_c$ are only given by bottom friction. 
    For the purpose of dimensional analysis, we will use here a simplified model for the friction force. The actual numerical method includes the full physics, as further discussed in Sec.~\ref{subsec:numericalapproach}. The friction force for a non-rotating, slipping particle is proportional to the normal force through the Coulomb friction coefficient $\mu_c$. Its explicit form is
    \begin{equation}\label{frictionforce}
       \mu_c\left|\boldsymbol{F}_c\cdot\hat{\boldsymbol{z}}\right|\hat{\boldsymbol{f}} = \mu_c\left|\left(\rho_s-\rho_f\right)V_s g-\boldsymbol{F}_L\cdot\hat{\boldsymbol{z}}\right|\hat{\boldsymbol{f}},
    \end{equation}
    where
    \begin{equation}\label{frictiondirection}
        \hat{\boldsymbol{f}} \equiv \frac{A\omega\sin(\omega t)\hat{\boldsymbol{y}}-\boldsymbol{u}_s}{\left|A\omega\sin(\omega t)\hat{\boldsymbol{y}}-\boldsymbol{u}_s\right|}
    \end{equation}
    is the unit vector that accounts for the relative velocity difference between the bottom plate and the particle. By definition, this vector always lies on the $xy$\nobreakdash-plane and is directed oppositely to the particle velocity relative to the bottom.

    Using the same typical scales as for the fluid, the dimensionless form for the integral of the stress tensor over the particle surface yields
    \begin{equation}
        \oint{\boldsymbol{\tau}\cdot\hat{\boldsymbol{n}}}dS = \rho_f\nu AD\omega\oint{\tilde{\boldsymbol{\tau}}\cdot\hat{\boldsymbol{n}}}{d\tilde{S}},
    \end{equation}
    where we have used that the surface area $\mathrm{S}$ of the spherical particle scales with $D^2$. The translation of the particles [Eq.~\eqref{particle_trans}] can then be rewritten in dimensionless form (skipping the tildes once again) as
    \begin{equation}\label{particle_trans_dimless}
        \frac{d\boldsymbol{u}_s}{dt} = 6\frac{\delta^{*2}}{s} \oint\boldsymbol{\tau}\cdot\hat{\boldsymbol{n}}dS + 2\pi\left(\frac{s-1}{s}\right)\frac{\mu_c}{\Gamma}\hat{\boldsymbol{f}}+\frac{2\pi}{s}\cos(2\pi t)\hat{\boldsymbol{y}}.
    \end{equation}
    Note that this equation now only considers the velocities in the $xy$\nobreakdash-plane, as opposed to the three-dimensional description in Eq.~\eqref{particle_trans}.
    The relative importance of the forcing terms in this equation can be varied independently from each other. The force caused by fluid-particle interaction (first term on the right-hand side) is proportional to the viscous drag and inversely proportional to the inertia of the particles relative to the fluid. The denser the particle with respect to the fluid and the larger the particle with respect to the Stokes boundary layer (over the sphere's surface), the smaller the acceleration due to the fluid-particle interaction is.
    For the friction with the bottom, we have neglected the contribution of lift forces, which are the result of complex particle-fluid interactions and thus not known a priori. Again, this is done only in this section, for the purpose of identifying dimensionless numbers. Subsequently, a combination of three dimensionless quantities ($s$, $\mu_c$, and $\Gamma$) sets the importance of bottom friction. Moreover, the second term on the right-hand side of Eq.~\eqref{particle_trans_dimless} is the only term where either $\Gamma$ or $\mu_c$ appear in the equations.
    
    Further note that there is no explicit dependence on $A^*$ in Eq.~\eqref{particle_trans_dimless} even though the drag force will likely implicitly depend on it. Because the particle velocity is scaled with $A\omega$, we implicitly assume that the excursion length of the particle is directly proportional to that of the fluid. In Sec.\ref{subsec:Quantitative}, we revisit the validity of this assumption. 
    
    The rotation of spherical particles in the system is described by 
    \begin{equation}\label{particle_rot}
        I_s\frac{d\boldsymbol{\omega}_s}{dt} = \oint\boldsymbol{r}\times\left(\boldsymbol{\tau}\cdot\hat{\boldsymbol{n}}\right)dS + \boldsymbol{T}_c,
    \end{equation}
    in which $\boldsymbol{\omega}_s$ is the angular velocity of the particle, $I_s=\rho_s\pi D^5/60$ is the moment of inertia of a solid sphere, $\boldsymbol{r}$ is the vector pointing from the sphere's center to its surface and $\boldsymbol{T}_c=\boldsymbol{r}\times \boldsymbol{F}_c$ is the torque exerted on the particle by collisions with either solid walls or other particles. Since the oscillating pressure gradient is uniform in space, it has no effect on the rotation of the particle, and we can exclude it from Eq.~\eqref{particle_rot}.
    Non-dimensionalization of this equation uses the dimensionless quantities introduced before in addition to
    \begin{equation}
        \boldsymbol{r}=D\tilde{\boldsymbol{r}}, \quad
        \boldsymbol{\omega}_s=\frac{A\omega}{D}\tilde{\boldsymbol{\omega}}_s, \quad 
        \oint\boldsymbol{r}\times\left(\boldsymbol{\tau}\cdot\hat{\boldsymbol{n}}\right)dS = \rho_f\nu AD^2\omega\oint\tilde{\boldsymbol{r}}\times\left(\tilde{\boldsymbol{\tau}}\cdot\hat{\boldsymbol{n}}\right)d\tilde{S}.
    \end{equation}
    Inserting these into Eq.~\eqref{particle_rot} and immediately dropping the tildes yields
    \begin{equation}\label{particle_rot_dimless}
        \frac{d\boldsymbol{\omega}_s}{dt} = 60\frac{\delta^{*2} }{s}\oint\boldsymbol{r}\times\left(\boldsymbol{\tau}\cdot\hat{\boldsymbol{n}}\right)dS + 20\pi\left(\frac{s-1}{s}\right)\frac{\mu_c}{\Gamma}\boldsymbol{r}\times\hat{\boldsymbol{f}},
    \end{equation}
    in which the same sets of dimensionless numbers as in Eq.~\eqref{particle_trans_dimless} appear. The first term on the right-hand side represents the contribution of the interaction with the fluid. Specifically, a spherical particle will rotate in a flow that has gradients in its velocity field such as shear flows or boundary-layer flows. The second term on the right-hand side gives the angular acceleration due to friction with the bottom and is zero if $\mu_c=0$.

\subsection{Simplifications}
    Due to our knowledge of the system, certain simplifications to the problem can be made \emph{a priori}. 
    First, the particles do not experience vertical motions and stay in contact with the bottom at all times. This means, as seen from Eqs.~\eqref{particle_trans_dimless} and \eqref{particle_rot_dimless}, that $\mu_c$ and $\Gamma$ do not need to be considered independently. Rather, only their ratio is relevant. This reduces the number of dimensionless quantities governing the problem to four. Moreover, when $\mu_c/\Gamma\ll1$, the translation of the particle is dominated by the pressure gradient and the hydrodynamic drag. In this case, the effect of varying $\mu_c/\Gamma$ on the translation of the particle is small, and the second term on the right hand side of Eq.~\eqref{particle_trans_dimless} can be neglected. However, weak bottom friction can still cause significant particle rotation because the uniform, oscillating pressure gradient does not directly affect the angular velocity. There is thus no term of order unity in Eq.~\eqref{particle_rot_dimless}, and hence the friction term is of leading order when it is larger than the torque induced by hydrodynamic drag and lift forces, i.e. when $(s-1)\mu_c/\Gamma\agt\delta^*$. Varying $\mu_c$ thus influences the type of motion of the particle (sliding vs. rolling), even when the effect of bottom friction on the particle excursion length remains small. 
    
    Second, the boundaries move in unison with the fluid in an oscillating box. This means that there is no vertical shear in the horizontal velocity above the bottom and thus also no Stokes boundary layer. This makes the system fundamentally different than the case of an oscillating flow over a fixed bottom. It also allows us to move to the reference frame in which both fluid and boundaries are at rest. Note that this is not an inertial frame because it undergoes a harmonically oscillating acceleration. The only motion in this frame is due to the particles moving relative to the fluid and the fluid that moves as a result of this relative particle motion. The absolute fluid excursion length $A^*$ has no intuitive physical meaning in this frame. Rather, we should consider the relative particle excursion with respect to the fluid $A_r^*$. This quantity is commonly referred to as the Keulegan-Carpenter number, and depends on the hydrodynamic drag, the solid friction, the inertia of the particle and is proportional to $A^*$ for constant values of $\delta^*$ and $s$ \cite{Wunenburger2002}. Hence the choice to scale $\boldsymbol{u}_s$ with $A\omega$ is still reasonable. At a fixed value of $A^*$, the value of $A_r^*$ can be varied between $0$ for $s=1$ and $A^*$ for $s\rightarrow\infty$.
    
    Finally, we consider the influence of $s$ on the system. This will be explicitly addressed later in Sec.~\ref{results_density} using simulation results. Here, we already show that despite the influence of $s$ on $A_r^*$, the density ratio does not influence the quasi-steady state of the system or therefrom derived quantities such as the mean gap between the particles.
    To show this, we set $\mu_c=0$ and convert Eq.~\eqref{particle_trans_dimless} to a power balance by taking the inner product with $\boldsymbol{u}_s$. Once the system has reached equilibrium, the kinetic energy of the particles when integrated over an integer number of periods $N$ does not change over time. This means that for the kinetic energy related to particle translation
    \begin{equation}\label{particle_power_balance}
        \int_{t}^{t+N}\frac{1}{2}\frac{d\left\|\boldsymbol{u}_s\cdot\boldsymbol{u}_s\right\|}{dt'}dt' = \int_{t}^{t+N}\left(6\frac{\delta^{*2}}{s} \oint\boldsymbol{\tau}\cdot\hat{\boldsymbol{n}}dS\cdot\boldsymbol{u}_s +\frac{2\pi}{s}\cos(2\pi t')\hat{\boldsymbol{y}}\cdot\boldsymbol{u}_s\right)dt'=0,
    \end{equation}
    and for the part related to particle rotation
    \begin{equation}\label{particle_power_balance_rot}
        \int_{t}^{t+N}\frac{1}{2}\frac{d\left\|\boldsymbol{\omega}_s\cdot\boldsymbol{\omega}_s\right\|}{dt'}dt' = \int_{t}^{t+N}\left(60\frac{\delta^{*2} }{s}\oint\boldsymbol{r}\times\left(\boldsymbol{\tau}\cdot\hat{\boldsymbol{n}}\right)dS\cdot\boldsymbol{\omega}_s \right)dt'=0.
    \end{equation}
    In both right-hand sides, $s$ can be eliminated. What remains is a balance between the mean power input and viscous dissipation in Eq.~\eqref{particle_power_balance}. This balance is governed explicitly by $\delta^*$ and implicitly by $A^*$ through $\boldsymbol{u}_s$, with both dimensionless numbers appearing also in the Navier-Stokes equation~\eqref{momentum_dimless} where they cannot be eliminated.
  
    In short, the density ratio has no direct effect on the mean state of the system, but does influence the relative amplitude of the particles. In turn, this may have an effect on the gap, such that the mean gap width is indirectly affected by $s$. In the simulations, we will first keep $s$ fixed and consider the influence of $A^*$ and $\delta^*$ on the mean state of the system. The value of $A_r^*$ changes with these numbers but is not set directly. As a validation, we explicitly look at the effects of varying $s$ in Sec.~\ref{results_density} and to the effects of non-zero bottom friction in Sec.~\ref{results_rotation}.
    
\section{\label{sec:Method}Numerical method}
    \subsection{Numerical approach}\label{subsec:numericalapproach}
    The numerical method used in this study is an updated version of the direct numerical simulation (DNS) and second-order accurate immersed boundary method (IBM) developed by \citeauthor{Breugem2012}~\cite{Breugem2012}, as recently used in \cite{Shajahan2020}. This method is in turn based on the work by \citeauthor{Uhlmann2005}~\cite{Uhlmann2005}. A more in-depth description and validation study can be found in the original sources. 
    
    In the DNS, the flow of a Newtonian, viscous fluid is solved on an equidistant, Cartesian grid that is fixed in space.
    On this so-called \textit{Eulerian} grid, the Navier-Stokes equation with the externally imposed pressure gradient is solved using a second-order accurate finite volume approach. A body force $\boldsymbol{F}_\mathrm{IBM}$ is added to ensure the no-slip/no-penetration condition at the particle surface by correcting the fluid velocity locally to the particle velocity.
    
    Each particle in the simulations is represented by a collection of points distributed over a spherical shell. The points on this \textit{Lagrangian} grid are fixed with respect to each other and the centroid of the particle. The position of the Lagrangian grid is free to change and is independent of the fixed Eulerian grid. The Lagrangian grid is not rotated due to the symmetry of its spherical shape. The translation of the particle's centroid and its angular velocity are solved using the Newton-Euler equations [Eqs.~\eqref{particle_trans_dimless} and \eqref{particle_rot_dimless}].
    
    A soft-sphere collision model is used for particle-wall and particle-particle interactions \cite{Costa2015}. Because the particles do not touch each other in our simulations, only the former is relevant here. The model further includes a stick/slip transition, accounts for (in)elastic deformation and rotation of the particles.
    In the direction normal to the collision (vertical in our case), a spring-damper model is used to model the force as 
    \begin{equation}
        \boldsymbol{F}_{c,n} = \left(k_n\delta_n + \eta_nu_n\right)\hat{\boldsymbol{z}}
    \end{equation}
    with the stiffness and damping coefficients $k_n$ and $\eta_n$. The spatial overlap and velocity difference between the particle and the bottom are given by $\delta_n$ and $u_n$, respectively \cite{Shajahan2020}.
    
    The tangential component is modelled as spring-damper interaction as well for the stick regime and as Coulomb friction in the slip regime
    \begin{equation}\label{ibmfriction}
        \boldsymbol{F}_{c,t} = \min\left(\left|k_t\delta_t+\eta_t u_t\right|, \mu_c \left|\boldsymbol{F}_{c,n}\right| \right)\hat{\boldsymbol{f}},
    \end{equation}
    with $\hat{\boldsymbol{f}}$ the direction of the force, oppositely to the (local) particle surface velocity relative to the bottom.
    Similar coefficients as for the normal components are used here, but now subscripts indicate tangential components. In all cases encountered in this study, the Coulomb friction term is the smallest of the two components of Eq.~\eqref{ibmfriction} during most of the particles' oscillations. Hence, we can use the collision term as it was already defined in Eqs.~\eqref{particle_trans_dimless} and \eqref{particle_rot_dimless}.
    
    The values of the coefficients $k_{n,t}$ and $\eta_{n,t}$ are calculated from the density ratio $s$, the dry coefficient of restitution $e_n=0.97$ and a tangential coefficient of restitution $e_t=0.39$. These values are based on experimental data and also used by \citeauthor{Shajahan2020}~\cite{Shajahan2020}. The collision time for particle-wall collisions is adapted to match the time step of the simulation. The full calculation of the coefficients is given in \cite{Costa2015}. 
    Since vertical particle motion is absent and friction is neglected for a major part of the results, the exact values of these coefficients are unlikely to be critical.
    
    Additionally, a lubrication model is used to resolve forces on particles at positions where the distance between the particle surface and bottom are smaller than an Eulerian grid cell \cite{Shajahan2020}. Details and validation of this model are given in \cite{Costa2015}.
    
    Both the Navier-Stokes and the Newton-Euler equations are integrated over time using an explicit three-step Runge-Kutta scheme \cite{Wesseling2001}. The time stepping is placed in a pressure-correction scheme.
    The time step $\Delta t$ for each simulation is such that it satisfies von Neumann stability \cite{Breugem2012}. Additional restrictions are added so that the total number of steps per oscillation period $1/\Delta t$ is an integer and that all output files are written after the same integer number of steps. This guarantees that the calculation of averaged quantities is always symmetric and includes exactly a full period.
    
    \subsection{Numerical setup}\label{subsec:numericalsetup}
    The resolution of the Eulerian grid size $\Delta x=1/16$ (with the particle diameter equal to unity). This ensures that flow structures can be resolved on a sub-particle level. This resolution is similar to the resolution used previously in other studies. For example, \citeauthor{Mazzuoli2016}~\cite{Mazzuoli2016} use 10-18 grid cells per particle diameter with a similar numerical method \cite{Kidanemariam2014}. \citeauthor{Klotsa2007}~\cite{Klotsa2007} use $13\frac{1}{3}$ grid cells per particle diameter. The resolution of the Lagrangian grid is similar to that of the Eulerian and consists of 746 points per particle.
    
    The simulations are set up as follows. The Eulerian grid covers a rectangular box with periodic boundary conditions in the horizontal directions. At the bottom and top, a no-slip/no-penetration condition is enforced. The plates are further set to oscillate with the same amplitude and frequency as the fluid. At height $z=5$, a stress-free boundary condition is enforced by setting the velocity gradient equal to zero. This acts as a symmetry plane and thus effectively halves the size of the computational domain. Formally, it also reflects the particles to the top plate. However, we have chosen $H^*$ sufficiently large such that the real and `virtual' particles do not significantly influence each other.
    For most simulations a length of $20$ and width of $15$ are chosen for directions parallel (streamwise) and perpendicular (spanwise) to the oscillation direction, respectively. The domain size is chosen to minimize the effects of the stress-free and periodic boundary conditions, while keeping the computational cost sufficiently low to allow for a scan of the parameter space. 
    More specifically, we have compared simulations using the different domain sizes $\left(20\times15\times5\right)$ and $\left(40\times20\times5\right)$. We found that the mean gap between the particles in quasi-steady state is affected less than 1\%. This difference is much smaller than other possible errors such as the extrapolation of the mean gap or the finite resolution of the spatial domain.
    
    At the start of each simulation, two particles are initialized on the bottom in a side-by-side configuration with respect to the oscillation direction. The initial distance between them is varied per simulation since it should ideally be close to the final (quasi)-steady configuration to save computational costs. The simulations are terminated once the pair has reached a steady state, which typically happens after approximately 20 to 50 periods, depending on the specific parameters. In some cases where the convergence is slow, the number of periods is extended to 100 to obtain a better estimate of the converged state of the pair.
    
    The used values of $\delta^*$ are $1/1.5, 1/2.25, 1/3.25, 1/4.5$ and $1/5.5$, given as reciprocals because, in the code, it is the value of $1/\delta^*$ that is actually set. The amplitude $A^*$ has values between approximately $0.37$ and $19.3$. The density ratio is kept constant at $s=7.5$ for the main results of this paper. This value is also used in previous experiments \cite{Klotsa2007,Klotsa2009}, and represents the ratio between stainless steel ($\rho\approx\SI{7950}{kg/m^3}$) and a 19:6 ratio mixture of water and glycerol ($\rho\approx\SI{1060}{kg/m^3}$).
    The particle-bottom interaction is controlled by fixing $\Gamma=4.5$ and varying the value of $\mu_c$. However, $\mu_c=0$ for the majority of the results. Only in Sec.~\ref{results_rotation}, the effect of particle rotation is studied through varying $\mu_c$.
    
    In Sec.~\ref{sec:Results}, the data from our simulations is compared to the experimental results by \citeauthor{Klotsa2007}~\cite{Klotsa2007}. The different types of markers used in their paper give the values of $D$, $\nu$ and $\omega$. From their figures, we further extract the values of $A_r$, $A_r\omega D/\nu$, $\Gamma$ and mean value of the gap $L$. This set is sufficient to reconstruct the dimensionless quantities $A^*$, $A_r^*$, $\delta^*$ and $L^*$. Only data for which all four numbers can be acquired are used. A file containing these data is available as supplementary material. Note that the simulations presented here span a larger region of the parameter space than the experiments, especially toward higher values of $A_r^*$ and $\delta^*$.
        
\section{\label{sec:Results}Results}
In this section, we present the results for our simulations of the particles in an oscillating box. First, a typical example of a particle trajectory is shown. In this example, the quantities of interest are identified and scaled as a function of flow conditions. After this, we directly compare with experimental results by \citeauthor{Klotsa2007}~\cite{Klotsa2007} to validate our simulations and show that our proposed scaling also holds for their data. Finally, average flow fields are used to understand the physics by coupling the quantitative data to the observed flow phenomena.

\subsection{Qualitative description of the particle trajectories}
\begin{figure*}
        \includegraphics{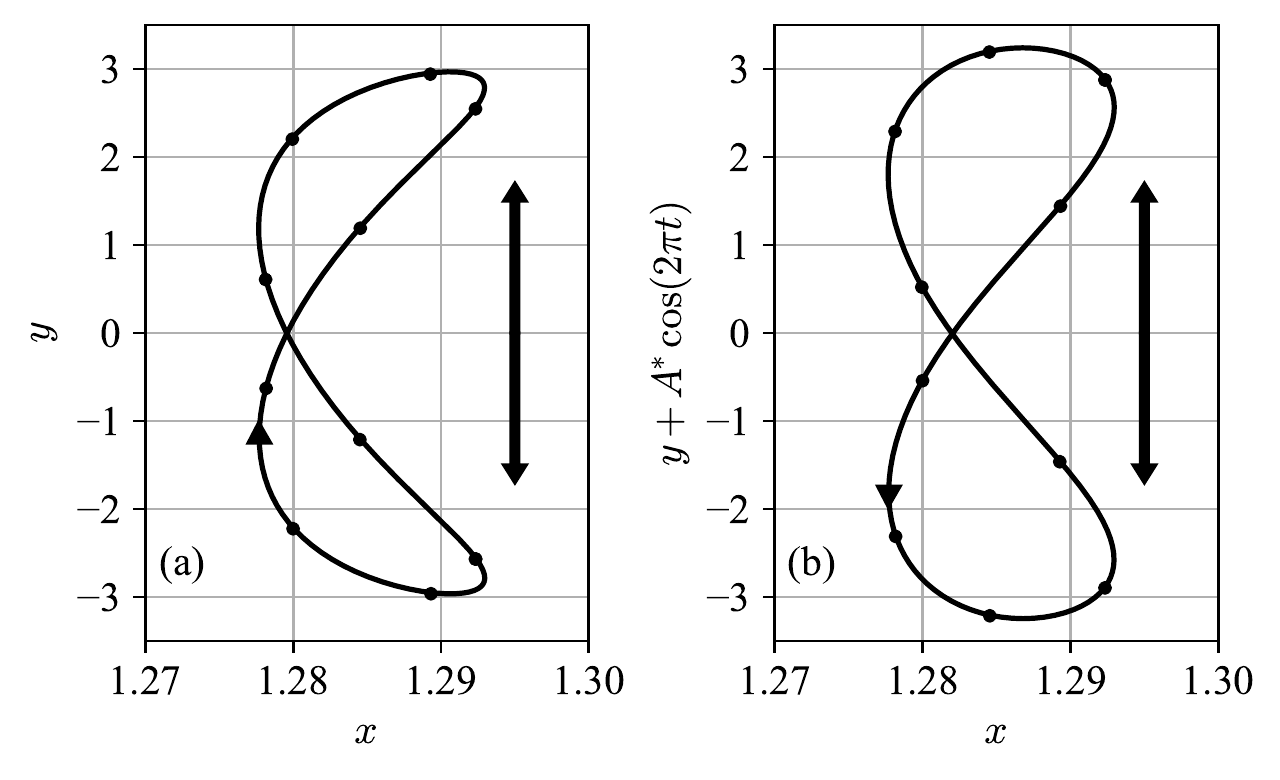}
        \caption{Trajectory of one of the particles forming a pair for one full period of the flow oscillation, as observed in the lab reference frame (a) and the reference frame moving with the fluid (b). The dots are drawn every 1/10\textsuperscript{th} of the oscillation period at identical times in both figures, so they can be directly compared. The arrows show the direction of motion. The axes are scaled such that both axes $x=0$ and $y=0$ are symmetry axes of the pair. The double-headed arrow indicates the direction of oscillation.
        } 
        \label{fig:ParticleTrajectory}
    \end{figure*}

    The motion of the center of one of the particles forming a pair is shown in Fig.~\ref{fig:ParticleTrajectory}, for a simulation with $A^*\approx 5.44$, $\delta^*=1/3.25$, $s=7.5$ and $\mu_c=0$. The trajectory shows the motion in the horizontal $xy$\nobreakdash-plane during one full oscillation period. To obtain the trajectory in the reference frame of the oscillating box, the fluid excursion in the $y$\nobreakdash-direction [given by $-A^*\cos(2\pi t)$] is subtracted from the particle position.
    Special care has been taken to show here a simulation for which the dynamics of the particle pair have reached a quasi-steady state. Once in this state, the pair remains stable, and the trajectory closes on itself after one period, i.e. the particle motion is periodic.
    Each trajectory is mirror symmetric in the $x$\nobreakdash-axis. Additionally, the trajectory of the particle that is not shown, is the mirror image with respect to the axis $x=0$ of the one in Fig.~\ref{fig:ParticleTrajectory}. The distance between the particle centers is thus approximately twice the value on the $x$-axis as shown here.
 
    The main motion of the pair is an oscillation in the direction of the driving flow (note the difference in scales between the two axes). The black arrows indicate the direction of the motion and are drawn at the same phase in both plots. The particle motion in the reference frame of the fluid is mirrored with respect to $y=0$ compared to its motion in the lab frame. This is because the particle lags behind the oscillating flow due to inertia, and thus, it moves `backwards' with respect to it.
    
    Perpendicular to the oscillating flow, we see an additional asymmetric oscillation at double the driving frequency $\omega$.
    The asymmetry of the trajectory (i.e. the skewness of the \emph{figure 8} shape in the $x$\nobreakdash-direction) indicates that higher harmonics are needed to describe its shape. 
    The amplitude of this oscillation is typically two orders of magnitude smaller than the main one. This might be the reason why this motion is reported here for the first time.
    We observe that the attraction between the particles coincides with high velocities relative to the fluid. They repel each other when their velocity is low, i.e. when they are approaching the extrema in the $y$\nobreakdash-direction in the `fluid frame'. As a result, the particle trajectories in a stable pair have characteristic \emph{figure 8} shapes.
    
\begin{figure*}
        \includegraphics{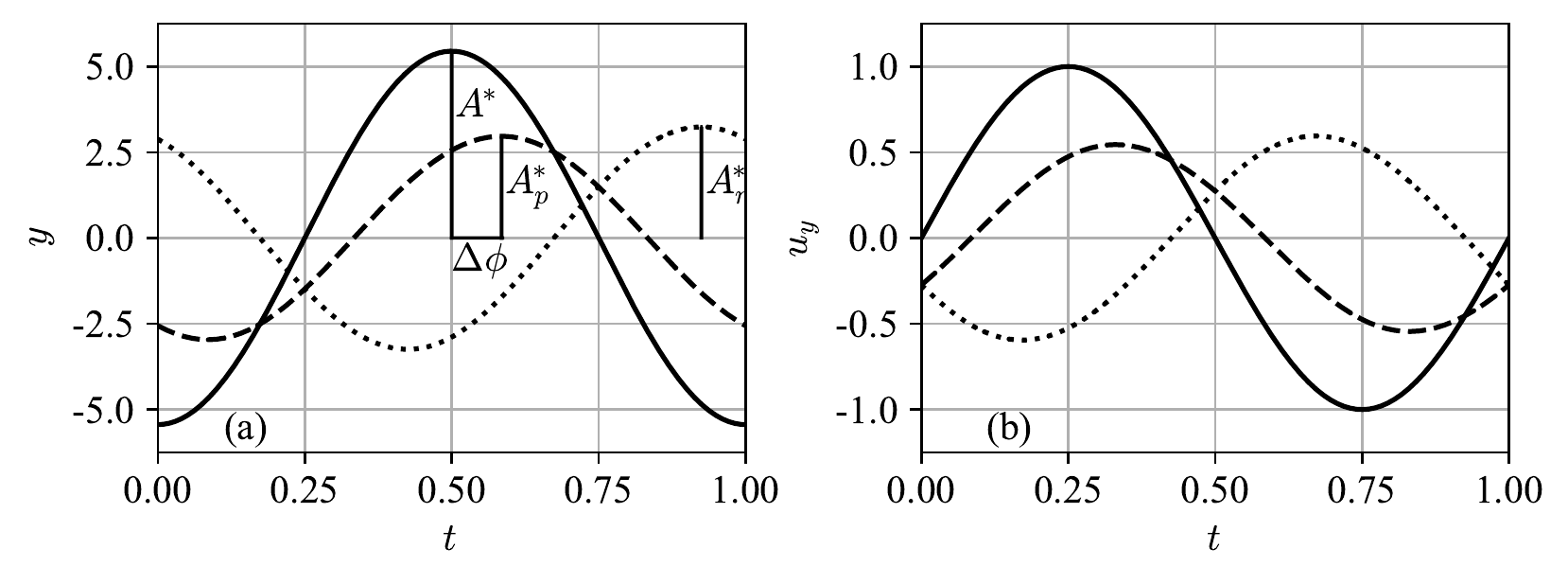}
        \caption{The position (a) and velocity (b) in the oscillation direction of one of the particles forming a pair is shown as a function of time during one period. The position is normalized using $D$, while the velocity and time are given in dimensionless form according to Eq.~\eqref{dimlesvars}. The different lines indicate the motion of the bulk flow and bottom (solid), or particle with respect to the lab frame (dashed) or fluid frame (dotted). The dimensionless amplitudes of the fluid, particle and relative motion ($A^*$, $A_p^*$ and $A_r^*$) and the phase difference ($\Delta\phi$) are labeled.}
        \label{fig:FrameComparison}
    \end{figure*}

    Figure~\ref{fig:FrameComparison} shows the position and velocity of the same particle as in Fig.~\ref{fig:ParticleTrajectory} during the same period. The normalized excursion length of the fluid is indicated with $A^*$. The normalized particle amplitude is $A_p^*$ in the lab frame and $A_r^*$ in the fluid frame of reference.
    Note that the values of $A_p^*$ and $A_r^*$ agree with the excursion lengths of the trajectory in the streamwise direction, as found in Fig.~\ref{fig:ParticleTrajectory}. Additionally, Fig.~\ref{fig:FrameComparison} shows that the direction of the particle motion reverses when switching to the fluid's frame of reference. 
    For both the position and velocity, there is a phase lag $\Delta\phi$ of the particle with respect to the bottom (i.e. the bulk of the fluid) due to inertia, i.e. $s>1$. The amplitudes and phase shift are related such that  
    \begin{equation}\label{amplitudes}
        A_r^* = \sqrt{{A^*}^2 + {A_p^*}^2 - 2A^*A_p^*\cos\left(\Delta\phi\right)},
    \end{equation}
    if we assume the curves to be perfect sinusoids. This expression simplifies to $A_r^*=\left|A^*-A_p^*\right|$ for $\Delta\phi=0$ and to $A_r^*=\sqrt{{A^*}^2 + {A_p^*}^2}$ for $\Delta\phi=\pi/2$. In general, $A_p^*$ is determined by Eq.~\eqref{particle_trans_dimless} and is thus a function of $A^*$, $\delta^*$ and $s$, in absence of friction.
    The phase shift $\Delta\phi$ is an unknown function of the same parameters. From Eq.~\eqref{amplitudes}, we learn that a scaling found for $A_p^*$ (possibly based on the governing equations) will generally not hold for $A_r^*$ due to the additional influence of this phase lag. Hence, we will use empirical data from the simulations to link $A_r^*$ to the flow conditions. The bulk flow velocity is subtracted from the particle velocity, after which a sinusoidal (as given in the Appendix~\ref{sec:Fitting}) is fitted through it. This method is equivalent to using Eq.~\eqref{amplitudes}.

    We make use of Figs.~\ref{fig:ParticleTrajectory}\nobreakdash-\ref{fig:FrameComparison} to introduce additional quantities that describe the particle pair. The quasi-steady state is given by the normalized, mean gap between the particles $L^*$. The dynamics within each period are characterized by the quantities describing the asymmetric oscillation of the gap size in the $x$\nobreakdash-direction. These quantities are the amplitudes of the multiple harmonics (2, 4 and 1 times the frequency $\omega$, in order of importance): $A_g^*=A_g/D$, $B_g^*=B_g/D$ and $C_g^*=C_g/D$. See the Appendix~\ref{sec:Fitting} for further details.
    
    The example in Fig.~\ref{fig:ParticleTrajectory} shows a particle trajectory that has reached a quasi-steady state (i.e. it no longer drifts over time). Simulations with lower values of $\mathrm{Re}$ or higher values of $\delta^*$ typically take hundreds of periods to reach this state. To avoid such long computations, the converged value of $L^*$ for these simulations is extrapolated using a least square fitting routine with the equations presented in the Appendix~\ref{sec:Fitting}. For each simulation, fits are used containing either an exponential decay or an inverse proportionality. Then, the values of the fit with the highest coefficient of determination $R^2$ are used.
    The fits for which we present derived quantities typically give $R^2>0.98$, with some exceptions for the highest values of $A^*$ where $R^2$ decreases to approximately $0.9$. Overall though, these values of $R^2$ give us confidence that the used equations describe the particle trajectories well. The difference in the extrapolated values of $L^*$ between the two fit variants is typically within a few ($<3\%$) percents of the value of $L^*$. The amplitudes of the harmonics are extracted using similar fitting equations, without extrapolation. These equations are further defined in the Appendix~\ref{sec:Fitting}.

\subsection{Quantitative description of the particle trajectories}\label{subsec:Quantitative}
\begin{figure}
        \includegraphics{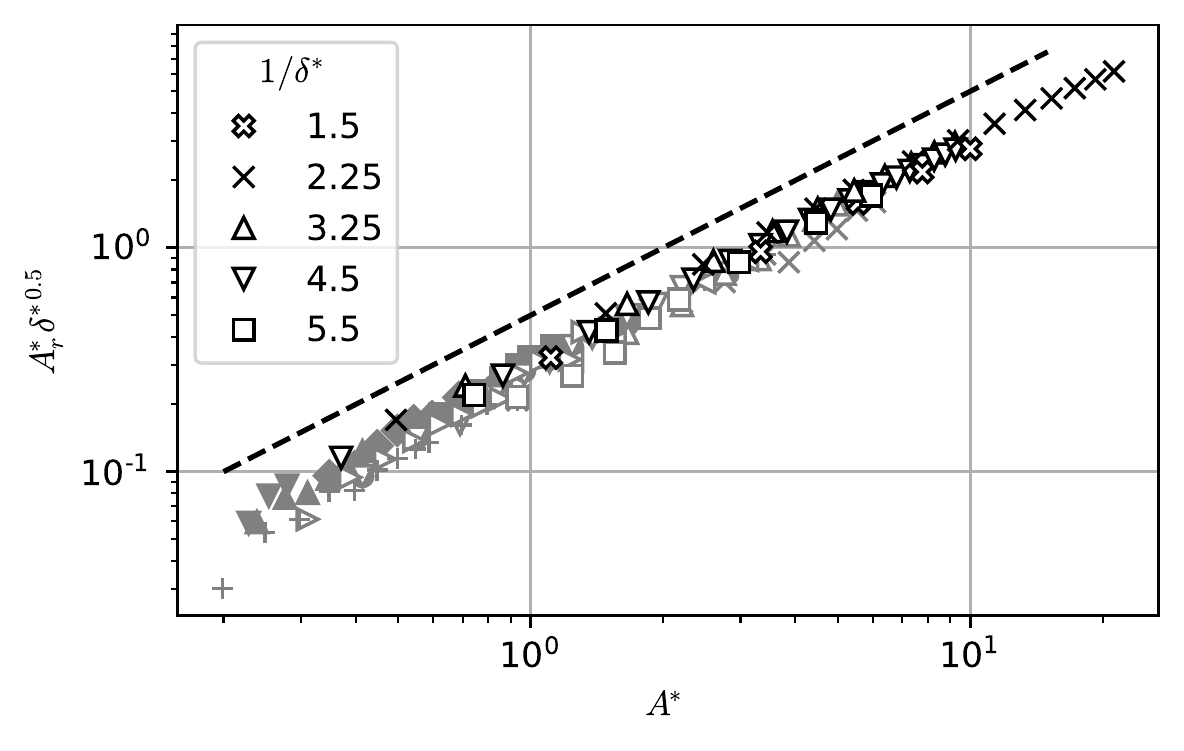}
        \caption{The particle amplitude relative to the fluid as a function of the absolute amplitude of the oscillating flow. Both data from our simulations (black) and experiments by \citeauthor{Klotsa2007}~\cite{Klotsa2007} (gray) are shown. A linear relationship is found between the two quantities, indicated by the dashed line with a slope of 1 dec/dec. Different markers refer to different values of $\delta^*$.}
        \label{fig:AD_Arel_osc}
    \end{figure}

    First, we consider the particle motion parallel to the driving flow. A linear relation is found between the relative amplitude $A_r^*$ and the oscillation amplitude in the lab frame $A^*$. 
    When $A_r^*$ is scaled with an additional factor ${\delta^*}^{0.5}$, the data for all simulations collapses onto a single line, as shown in Fig.~\ref{fig:AD_Arel_osc}. The linear relation between $A_r^*$ and $A^*$ (indicated by the dashed line) is expected when considering Eq.~\eqref{particle_trans_dimless}, where the particle velocity is scaled with $A\omega$. The proportionality constant between $A_r^*$ and $A^*$ depends on the viscous length scale and thus on $\delta^*$ (indicated by different markers). Moreover, the experimental data by \citeauthor{Klotsa2007}~\cite{Klotsa2007}, shown in the same figure, collapses onto the same curve.
    
    From Fig.~\ref{fig:AD_Arel_osc}, we obtain the scaling
    \begin{equation}\label{eq:Ar}
        A_r^* = C\frac{A^*}{{\delta^*}^{0.5}},
    \end{equation}
    where $C\approx0.30$ is a constant that depends on $s$ and $\mu_c/\Gamma$ according to Eq.~\eqref{particle_trans_dimless}. Equation~\eqref{eq:Ar} implies that for higher $\delta^*$ values, both viscous effects and the magnitude of the particle-fluid interaction increase. The particle then moves more along with the fluid, and hence, it has a smaller relative amplitude.
    For the region of the parameter space considered in our simulations, the linear relation between $A_r^*$ and $A^*$ holds. However, Eq.~\eqref{eq:Ar} cannot hold when considering strongly nonlinear drag or turbulent flows, see e.g. \cite{Martin1976}.
    Nevertheless, a linear dependence of $A_r^*$ on $A^*$ was previously observed in experiments \cite{Wunenburger2002,Klotsa2007}. Using a phenomenological model, \citeauthor{Wunenburger2002}~\cite{Wunenburger2002} showed that the proportionality between the amplitudes is governed by the viscous drag force, particle inertia and bottom friction. 
    Since we have chosen $\mu_c=0$ and $s=7.5$ as constants in our simulations, only the effect of a varying viscous drag force is seen. Hence, we only find a dependence on $\delta^*$ in Eq.~\eqref{eq:Ar}.
    Moreover, we identified in Eq.~\eqref{particle_trans_dimless} that when $s$ is constant and $\mu_c=0$, the magnitude of the particle-fluid interaction is governed by $\delta^*$.

\begin{figure}
        \includegraphics{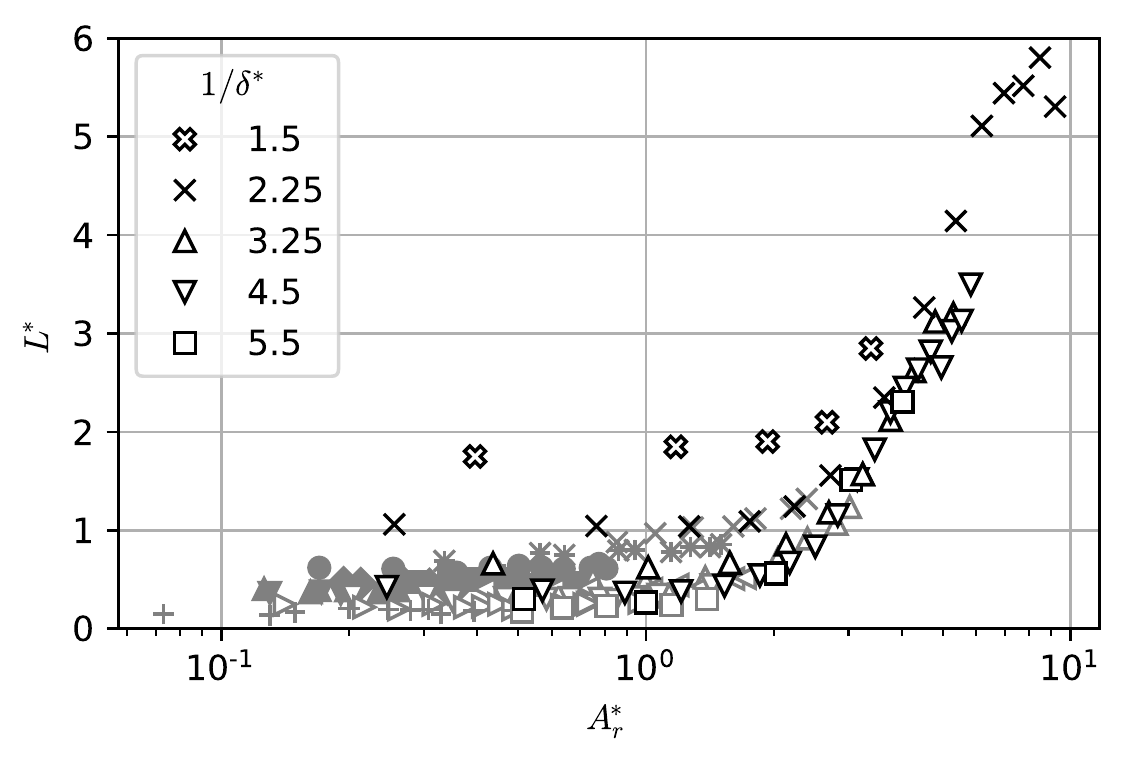}
        \caption{The mean value of the gap as a function of the relative amplitude of the particles for both our simulations (black) and experiments by \citeauthor{Klotsa2007}~\cite{Klotsa2007} (gray). A viscous and an advective regime are found, with a transition around $1\alt A_r^*\alt 2$. Different markers refer to different values of $\delta^*$.}
        \label{fig:sD_Arel_osc}
    \end{figure}

    We now consider the mean value of the gap $L^*$ as a function of the relative amplitude $A_r^*$ as shown in Fig.~\ref{fig:sD_Arel_osc}. The horizontal axis is logarithmic for the sake of clarity. Each set of simulations for a constant $\delta^*$, indicated with different markers, shows a strikingly similar behavior. For $A_r^*\alt 1$, the distance between the particles depends only on the viscous length scale $\delta^*$ (i.e. it is independent of $A_r^*$). Hence, we refer to this region of the parameter space as the `viscous regime'. For $A_r^*\agt 1$, the mean distance between the particles increases superlinearly with $A_r^*$. We refer to this region of the parameter space as the `advective regime'.
    
    There is a good quantitative agreement between our simulations and the experimental data by \citeauthor{Klotsa2007}~\cite{Klotsa2007}. In particular for similar $\delta^*$ values ($\delta^*=1/2.25, 1/3.25, 1/5.5$ denoted by crosses, upward triangles and squares respectively), the typical difference in the value of $L^*$ between simulations and experiments is only $10\%$. This is significantly more accurate than the simulation results in the original article \cite{Klotsa2007}, which overestimated the gap by approximately a factor of two.
    
\begin{figure}
        \includegraphics{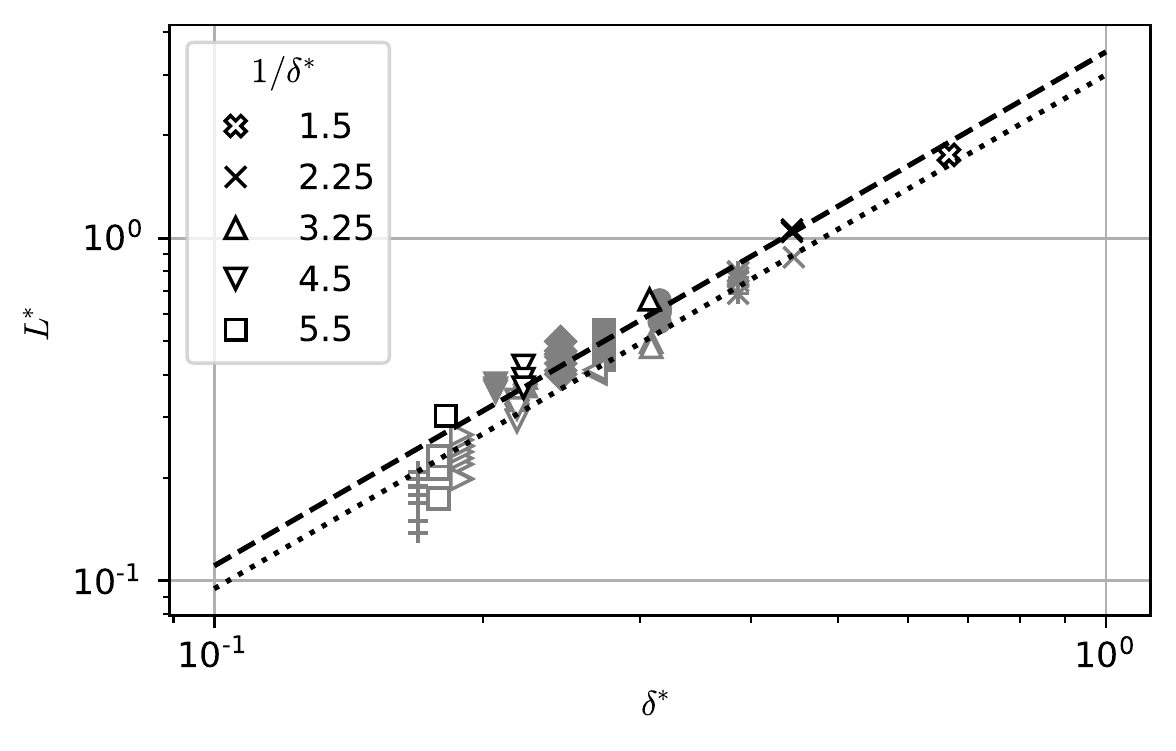}
        \caption{Scaling of the data with $A_r^*\leq1.0$ for data from our numerical simulations (black) with the experimental data from \citeauthor{Klotsa2007}~\cite{Klotsa2007} (gray). The dotted and dashed lines indicate the lines $L^*=3.0{\delta^*}^{1.5}$ and $L^*=3.5{\delta^*}^{1.5}$, respectively.}
        \label{fig:sD_delta_comp}
    \end{figure}

    We now focus on the viscous regime by plotting only the markers with $A_r^*\leq1.0$ as a function of $\delta^*$ in Fig.~\ref{fig:sD_delta_comp}. As can be seen, the mean value of the gap scales with ${\delta^*}^{1.5}$. We use only our simulation data to obtain this scaling because we have a fuller understanding of its accuracy compared to that of the experimental data. Nevertheless, the scaling holds for most experimental results as well. The discrepancy at lower values of $\delta^*$ are magnified due to the logarithmic scale on the vertical axis. The magnitude of the deviations falls within the typical uncertainty of around $0.1D$.
    
    Determining the proportionality constant between $L^*$ and ${\delta^*}^{1.5}$ based on Fig.~\ref{fig:sD_delta_comp} is more ambiguous. If for each $\delta^*$ value, only the minimum value of $L^*$ is used, a constant of approximately $3.0$ is found. This value also matches best with the minima of the experimental data. Using all shown (black) markers yields a constant of about $3.5$. In short, for the viscous regime, we obtain the scaling
    \begin{equation}\label{eq:scaling_viscous}
        L^*\approx\alpha {\delta^*}^{1.5},
    \end{equation}
    with $3.0 \alt \alpha \alt 3.5$. From here on, we will use $\alpha=3.0$ as this gives a better collapse of $L^*$ also in the advective regime, which will be addressed next.
    
    We now collapse the data from Fig.~\ref{fig:sD_Arel_osc} onto a single curve. This is commonly done by division, but here we subtract Eq.~\eqref{eq:scaling_viscous} from the value of $L^*$. This approach also collapses the values for $A_r^*\agt 1$. Even more important, both the simulation and experimental data collapse onto the same curve, showing a smooth transition between the two regimes at $1\alt A_r^*\alt 2$. We find no dependence on $\delta^*$ for the transition location. This result is different than the previously reported $\delta^*$-dependent transition at $A_r^*\approx 7\sqrt{\nu/\omega}/D\approx 5\delta^*$ \cite{Klotsa2007}. 
        
\begin{figure}
        \includegraphics{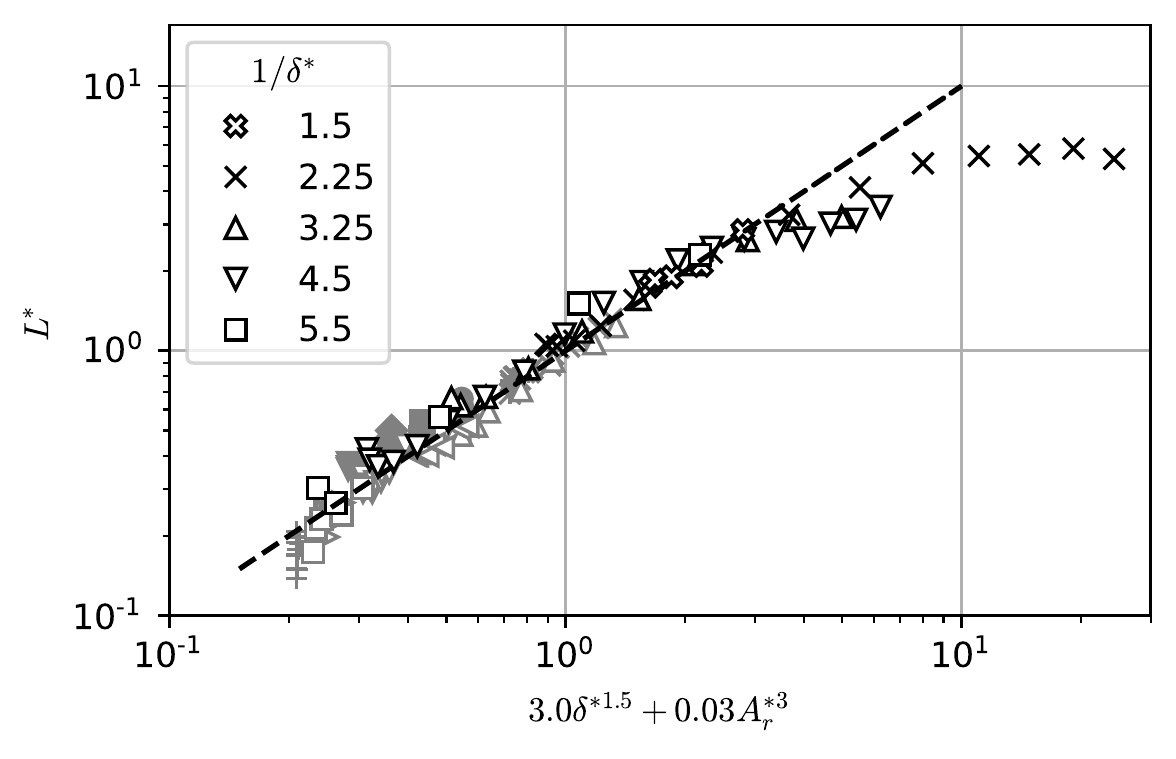}
        \caption{Scaling of both regimes for data from both our simulations (black) and experiments by \citeauthor{Klotsa2007}~\cite{Klotsa2007} (gray). The dashed curve indicates the identity line $L^*=3.0{\delta^*}^{1.5} + 0.03 {A_r^*}^3$.}
        \label{fig:sD_Arel_collapse}
    \end{figure}

    Figure~\ref{fig:sD_Arel_collapse} shows that the mean gap scales overall according to 
    \begin{equation}\label{scalinggap}
        L^* \approx 3.0{\delta^*}^{1.5} + 0.03 {A_r^*}^3.
    \end{equation}
    This scaling means that, when the first term dominates over the second term, viscous forces dominate and the mean size of the gap is only determined by the viscous length scale: the Stokes boundary layer thickness. This happens when $A_r^*\ll 4.6{\delta^*}^{0.5}$. We may additionally substitute Eq.~\eqref{eq:Ar} to find $A/\delta \ll 4.6/C(s)$, but this gives a transition that is highly dependent on $s$.
    For $2 \alt A_r^*\alt 4$, non-linear effects make the gap grow with $0.03 {A_r^*}^3$. This superlinear scaling holds for a rather short range, so the exact values for the exponent and proportionality constant should be considered with care. Still, it is clear that the gap size increases sharply with $A_r^*$ in this regime.
     
    For the highest values of $A_r^*$ obtained in the simulations, the gap is notably smaller than predicted by the proposed scaling. This is especially clear for the five crosses on the top right of Fig.~\ref{fig:sD_Arel_collapse}. In these simulations, the mean spacing of the gap is comparable to half the domain size. As a result, the instantaneous interaction between the particles is weaker, as will be further addressed when the oscillation of the gap is discussed. The weakening interaction allows other mechanisms to influence the particle pair configuration and its stability. 
    One possibility is that particles interact through the periodic boundaries. However, increasing the domain size by a factor two yields a difference of only 1\% in the value of $L^*$.
    An increase in the spatial resolution to $\Delta x=1/24$ at high values of $A_r^*$ does yield larger $L^*$ values that lie closer to the proposed scaling. High resolution simulations in the viscous regime yield no significant deviations in the value of the mean gap.
    We believe that further high-resolution simulations for large $A_r^*$-values do not provide added value for this study. Furthermore, there is no experimental data of this regime currently available to compare with the simulations. Most probably, pair formation cannot even be observed experimentally for these large $A_r$ values due to the weak particle-particle interactions at such large distances.
    
    Even though Eq.~\eqref{scalinggap} accurately describes the mean spacing of the gap, it uses two independent quantities ($A_r^*$ and $\delta^*$) to describe a single curve. This suggests there is a single underlying quantity that describes the two regimes and the transition between them. This quantity emerges when considering the Reynolds number of the mean gap, given by 
    \begin{eqnarray}\label{ReL}
        \mathrm{Re}_L = \frac{A_r\omega L}{\nu} &=& \frac{2A_r^*L^*}{{\delta^*}^2} \\
        &\approx& 6\frac{A_r^*}{{\delta^*}^{0.5}}\left[1+0.01\left(\frac{A_r^*}{{\delta^*}^{0.5}}\right)^3\right] \nonumber \\
        &\approx& 6C(s)\frac{A}{\delta}\left[1+0.01\left(C(s)\frac{A}{\delta}\right)^3\right] \nonumber
    \end{eqnarray}
    where the latter two equations are obtained by filling in the previously found scalings from Eqs.~\eqref{eq:Ar}~and~\eqref{scalinggap}. The expression on the 2\textsuperscript{nd} line is a function of the only two parameters governing the quasi-steady state of the system. It has a form such that effectively there is only one degree of freedom, namely $A_r^*/{\delta^*}^{0.5}$. The expression on the 3\textsuperscript{rd} line is only a function of the control parameters of the problem and shows the ratio $A/\delta$ which appeared first in Eq.~\eqref{momentum_dimless_alt} and was interpreted as the Reynolds number of the Stokes boundary layer in Eq.~\eqref{reynoldsdelta}.
    
    No explicit information about the transition is included in the definition of $\mathrm{Re}_L$ or in the rest of the derivation. Still, Eq.~\eqref{ReL} has a smooth transition between two drastically different regimes: $\mathrm{Re}_L \approx 6A_r^*/{\delta^*}^{0.5}$ in the viscous regime and $\mathrm{Re}_L \approx 0.06\left(A_r^*/{\delta^*}^{0.5}\right)^4$ in the advective regime.
    The transition between these two regimes starts approximately when the second term between the square brackets becomes significant ($\approx 10\%$) with respect to the first term, i.e. $0.01\left(A_r^*/{\delta^*}^{0.5}\right)^3\approx 0.1$. This yields a transition criterion of $A_r^*\approx 10^{1/3}{\delta^*}^{0.5}$. For our simulation, this results in values of $A_r^*\approx 1.8$ for $\delta^*=1/1.5$ and $A_r^*\approx 0.9$ for $\delta^*=1/5.5$. 
    These outcomes match with the transition region at $1\alt A_r^*\alt 2$ found previously in Fig.~\ref{fig:sD_Arel_osc}, but are not independent of $\delta^*$. An explanation is that the range of $\delta^*$ values considered here is too small, such that the transition seems to be at constant $A_r^*$. It is especially difficult to notice small changes from the discrete set of markers in e.g. Fig.~\ref{fig:sD_Arel_osc}. Revisiting Fig.~\ref{fig:sD_Arel_osc}, it indeed seems that the squares transition at lower $A_r^*$ than the (open) crosses.

    
    We now describe the oscillation in the direction perpendicular to the oscillatory flow. The values of the amplitudes $A_g^*$ and $B_g^*$ of the oscillation of the gap, as defined in the Appendix~\ref{sec:Fitting}, are shown in Figs.~\ref{fig:Agap_Arel_osc} and \ref{fig:Bgap_Arel_osc}, respectively. With $A_r^*$ on the horizontal axis and the amplitudes scaled with powers of $\delta^*$, the data collapses on a curve. A sharp change in behavior is seen around $A_r^*\approx2$, which is the same value as for the start of the ${A_r^*}^3$ increase in the mean gap $L^*$. The consistency of the $A_r^*$-dependent transition for $L^*$, $A_g^*$ and $B_g^*$ gives us confidence that this is an improvement over the previously proposed transition $A_r^*\approx5\delta^*$ \cite{Klotsa2007}.
    For $A_r^*\alt 2$, 
    \begin{subequations}
    \begin{equation}
        A_g^*{\delta^*}^2 \propto {A_r^*}^2,
    \end{equation}
    \begin{equation}
        B_g^*{\delta^*}^{2.5} \propto  {A_r^*}^4.
    \end{equation}
    \end{subequations}
    These expressions can be rewritten as
    \begin{subequations}\label{AB_osc_scaling}
    \begin{equation}
        A_g^* = C_A\left(\frac{A_r}{\delta}\right)^2 = C_A'\left(\frac{A}{\delta}\right)^2\frac{1}{\delta^*},
    \end{equation}
    \begin{equation}
        B_g^* = C_B \left(\frac{A_r}{\delta}\right)^{2.5}{A_r^*}^{1.5}=C_B'\left(\frac{A}{\delta}\right)^4\frac{1}{{\delta^*}^{0.5}},
    \end{equation}
    \end{subequations}
    with $C_A\approx 5\times 10^{-4}$ and $C_B\approx 4\times 10^{-6}$ both constants.
    This means that both the oscillation amplitude (through $A_g^*$) and the asymmetry of this oscillation (through $B_g^*$) increase with the relative particle amplitude and decrease when viscous effects become more important. 
    
    For $A_r^*\agt 2$, the amplitude $A_g^*$ decreases and the data shows more scatter. This decrease is steeper for lower values of $\delta^*$, when viscous damping effects are less important. The value of $B_g^*$ decreases less drastically after the transition and stays within the range $1-5\times 10^{-5}$ for simulations considered here.
    The decrease of both amplitudes in the advective regime, when $L^*$ becomes large, implies that the instantaneous interactions between the particles gets weaker once they are further apart. This allows other mechanisms to influence the particle pair and their stability. In an experiment, an observer could see the gap becoming unstable due to influence of e.g. boundaries, surface imperfections or slight non-uniformity in the flow. So all in all, the interpretation of the markers at the highest values of $A_r^*$ should be considered with caution.

    No clear scaling or collapse was found for $C_g^*$. This is to be expected because the fluid oscillates symmetrically, and hence, there should not be a preferential direction once transients from the startup have decayed.

    \begin{figure}
        \includegraphics{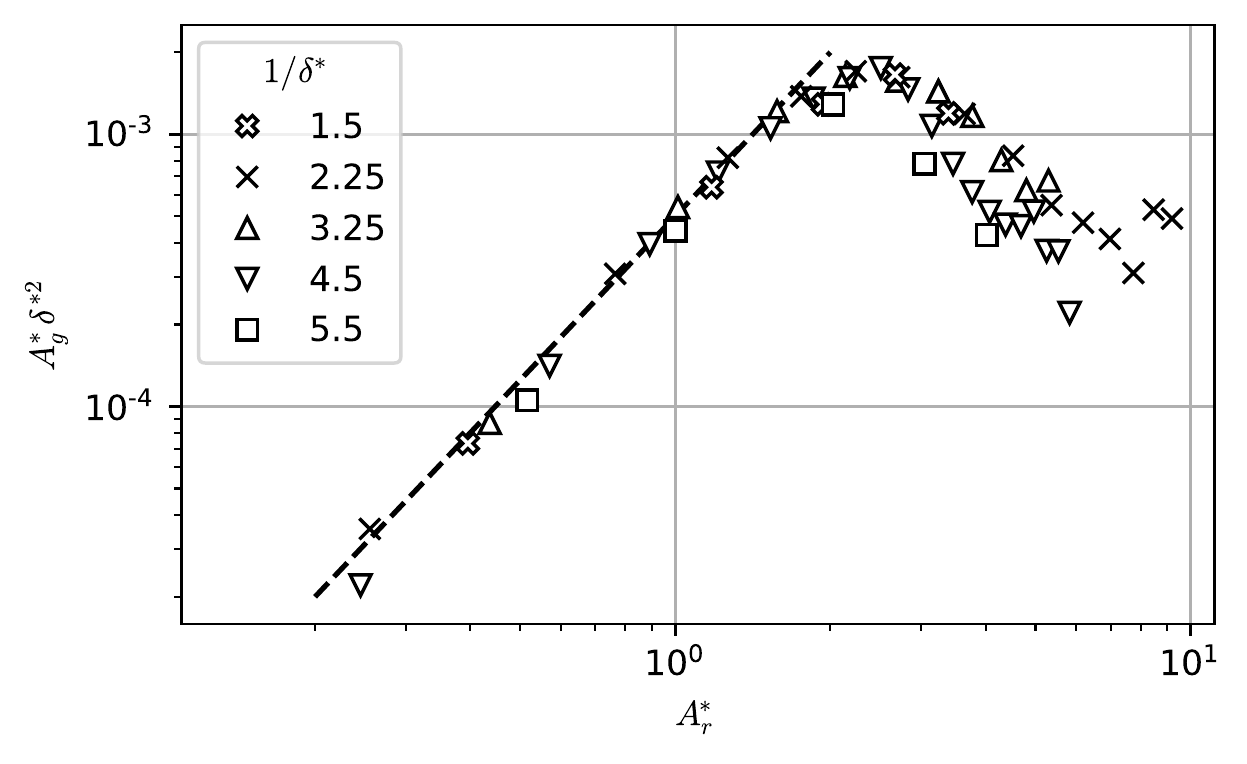}
        \caption{The amplitudes of the gap oscillation at two times the driving frequency as a function of the relative particle amplitude with respect to the fluid. Two distinct regimes are found, with a transition around $A_r^*\approx2$. The dashed line has a slope of 2 dec/dec. Different markers refer to different values of $\delta^*$.}
        \label{fig:Agap_Arel_osc}
    \end{figure}
    \begin{figure}
        \includegraphics{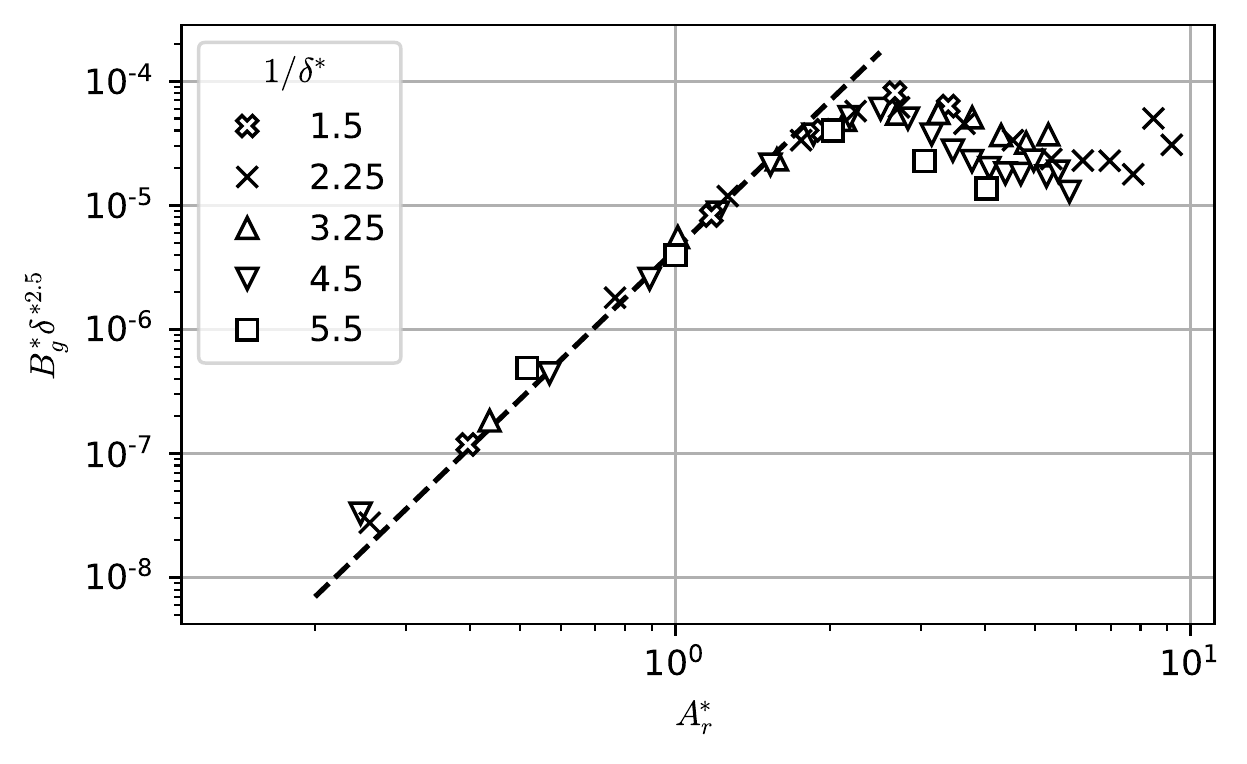}
        \caption{The amplitudes of the gap oscillation at four times the driving frequency as a function of the relative particle amplitude with respect to the fluid. Two distinct regimes are found, with a transition around $A_r^*\approx2$. The dashed line has a slope of 4 dec/dec. Different markers refer to different values of $\delta^*$.}            
        \label{fig:Bgap_Arel_osc}
    \end{figure}

\subsection{Steady streaming flows}
\begin{figure*}
        \includegraphics{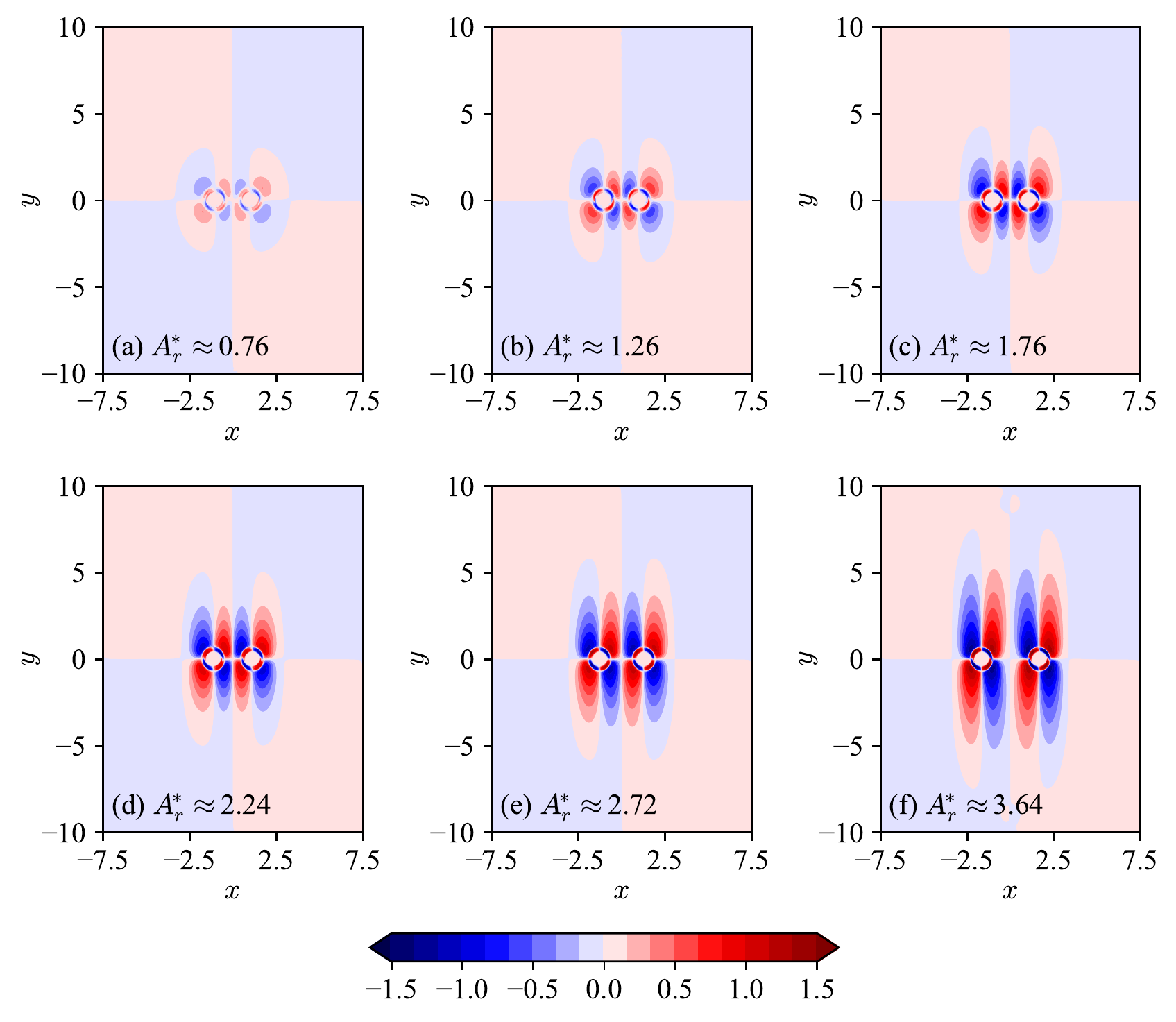}
        \caption{(Color online) Averaged vorticity fields in the $xy$-plane going through the center of the particles for simulations with $\delta^*=1/2.25$ and different values of $A_r^*$. The quantity $\text{sign}(\left<\omega_z\right>)\log\left(\left|\left<\omega_z\right>\right|+1\right)$ is plotted for visualization purposes, where negative vorticity (blue) leads to a clockwise flow and positive vorticity (red) to a counterclockwise flow. The light circular patches indicate the positions of the particles.}
        \label{fig:steadystreamingvort}
    \end{figure*}

    We will now consider the flow fields around the particle pairs and show how they are affected by changes in the flow parameters. In Fig.~\ref{fig:steadystreamingvort}, the average $z$-component of the vorticity field in the horizontal plane going through the center of the particles is shown for different values of $A_r^*$. It is defined as 
    \begin{equation}\label{eq:vorticity_definition}
        \left<\omega_z\right> = \frac{1}{N}\int_{t}^{t+N} \left(\frac{\partial u_y}{\partial x}-\frac{\partial u_x}{\partial y}\right)dt'.
    \end{equation}
    Due to the finite number of time steps in the simulations, the integral is replaced by a summation over a number of time frames. In our case, each output file is written after exactly 1/20\textsuperscript{th} of the oscillation period, i.e. at (dimensionless) times $t, t+1/20,\dots,t+N-1/20,t+N$.
    We calculate Eq.~\eqref{eq:vorticity_definition} from the reference frame moving with the particles, such that Fig.~\ref{fig:steadystreamingvort} shows the flow relative to the pair. The six subplots can be roughly divided into three in the viscous regime and three in the advective regime.
    
    Vorticity is generated at the no-slip boundaries of the spheres, which is here visible by the thin shell directly surrounding the particles. Each half period, part of the vorticity gets advected from the particle by the driving flow. On average, there is a symmetric distribution of `outer' vorticity patches around the pair, which are larger in space but weaker in strength than the vorticity concentrated in the thin shell. This is especially clear at the lower relative amplitudes in Fig.~\ref{fig:steadystreamingvort}.

    One should note that the vorticity patches shown here are actually cross sections of vortex rings. In the absence of a bottom plate, closed loops would form. For the case of a single, isolated particle, these are rotationally symmetric around the $y$\nobreakdash-axis going through the center of the sphere.
    The solid bottom breaks the symmetry, and half rings, that terminate at the bottom, form instead \cite{Klotsa2009}.
    
\begin{figure*}
        \includegraphics{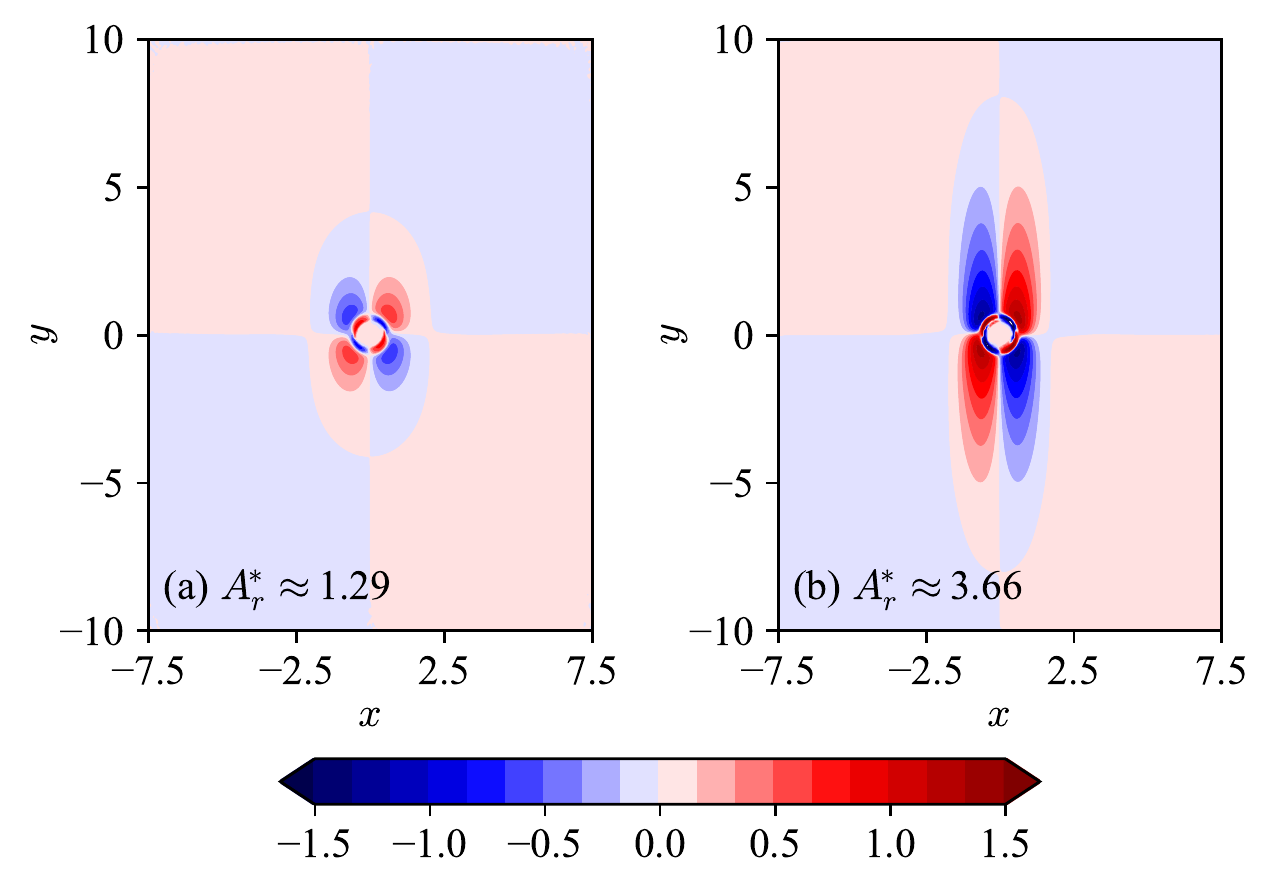}
        \caption{(Color online) Averaged vorticity fields in the $xy$-plane going through the center of a single particle for simulations with $\delta^*=1/2.25$ and values of $A_r^*$ close to those presented in Fig.~\ref{fig:steadystreamingvort}. The quantity $\text{sign}(\left<\omega_z\right>)\log\left(\left|\left<\omega_z\right>\right|+1\right)$ is plotted for visualization purposes, where negative vorticity (blue) leads to a clockwise flow and positive vorticity (red) to a counterclockwise flow.}
        \label{fig:steadystreamingvort1p}
    \end{figure*}

    In the viscous regime, the mean spacing between the spheres is constant with respect to $A_r^*$, while the magnitude of the vorticity around them does increase with $A_r^*$. The structure of the vortex patches in this regime is always similar. The patches between the particles are narrower ($x$-direction) and slightly shorter ($y$-direction) than those on the outer sides of the pair. Only the part of the flow field on the outside of the pair resembles that around a single particle as can be seen when comparing with Fig.~\ref{fig:steadystreamingvort1p}. Qualitatively similar steady streaming flows are found in numerical simulations by \citeauthor{Jalal2016}~\cite{Jalal2016} for fixed spheres in the limit of $A_r^*\rightarrow 0$.
    
    When the value of $A_r^*$ increases across the transition between regimes, the magnitude of the vorticity patches continues to increase with it. All outer vorticity patches become more elongated, with a length of approximately $2A_r^*$ in the streamwise direction. This agrees with literature in which the relative wake length behind an obstacle in oscillating flow scales with the Keulegan-Carpenter number \cite{Branson2019}, which is equivalent to $A_r^*$.
    The vorticity distribution around each particle now resembles that of the single, isolated particle, as shown in Fig.~\ref{fig:steadystreamingvort1p}. Additionally, it is noticeable that the vortices between the pair become slightly longer than the outer vortices around the transition $A_r^*\approx2$.
    
\begin{figure*}
        \includegraphics{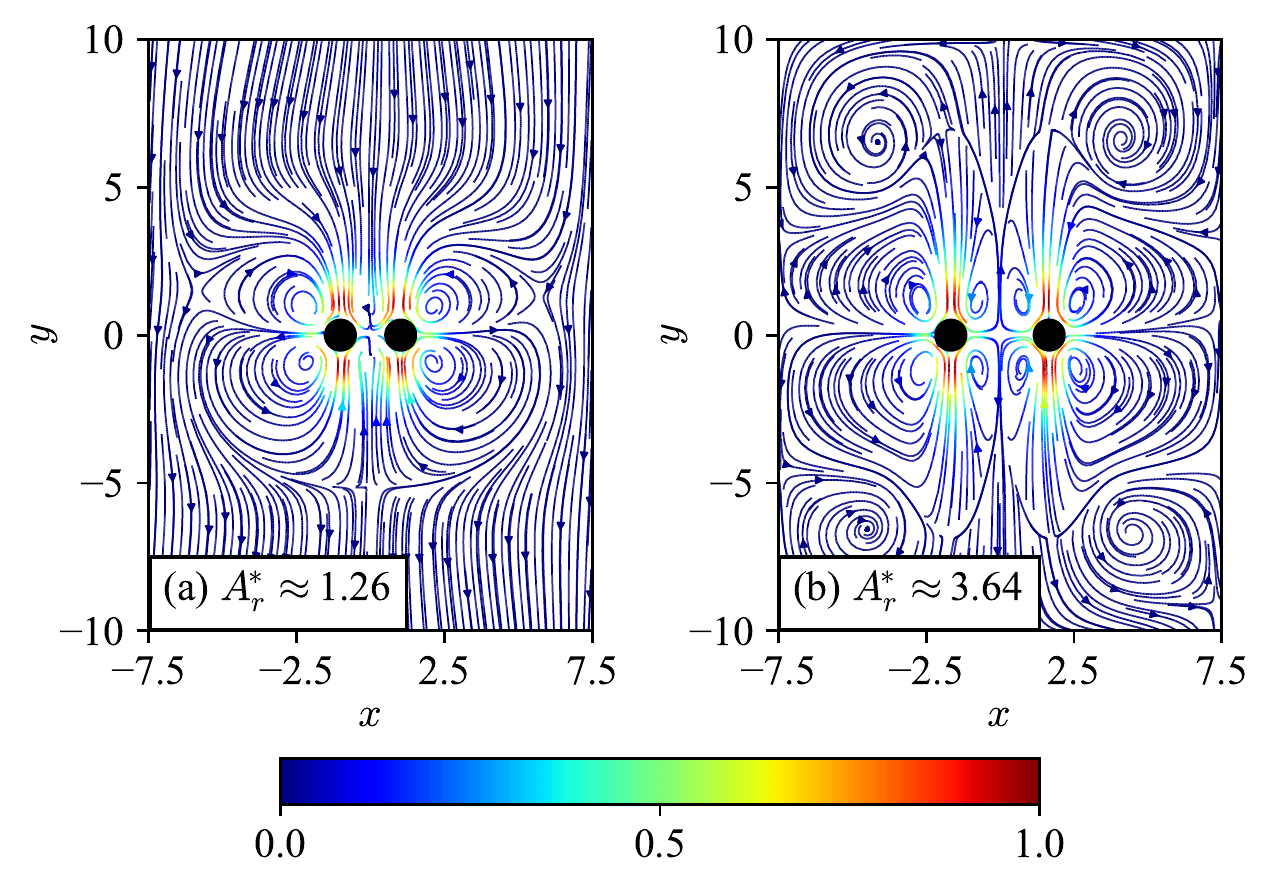}
        \caption{(Color online) The streamlines of the averaged velocity field around the particle pair in the $xy$-plane going through the particle centers. Two simulations are shown with $\delta=1/2.25$, corresponding to the viscous (a) and advective (b) regime. The color of the lines corresponds to the magnitude of the velocity field, normalized by its maximum value. The mean positions of the particles are given by the black circles.}
        \label{fig:steadystreamingsteamlines}
    \end{figure*}

    The streamlines of the averaged velocity field are shown in Fig.~\ref{fig:steadystreamingsteamlines} for two of the simulations from Fig.~\ref{fig:steadystreamingvort}. Between the particles the structure of the steady streaming flow changes smoothly with $A_r^*$. In the viscous regime, in Fig.~\ref{fig:steadystreamingsteamlines}a, the flow is dominated by a `jet'-like flow away from the pair in the $y$-direction.
    In the advective regime, in Fig.~\ref{fig:steadystreamingsteamlines}b, the `jet'-like flow transitions into a pair of vortices that grow in size as the gap between the particles increases. Simultaneously, a secondary set of four vortices appears further away from the pair, filling the outer parts of the domain. 
    These two streaming flow regimes were also found in experiments and simulations by \citeauthor{Klotsa2007}~\cite{Klotsa2007}. 

\subsection{Effect of density variation}\label{results_density}
    So far, only results from simulations with density ratio $s=7.5$ have been presented. We now present results for $s=2.65$. This value is common for natural sediment and was previously used in simulations by \citeauthor{Mazzuoli2016}~\cite{Mazzuoli2016}. We only performed simulations for $\delta^*=1/4.5$ and $\mu_c=0$ while varying $A^*$.
    
\begin{figure}
        \includegraphics{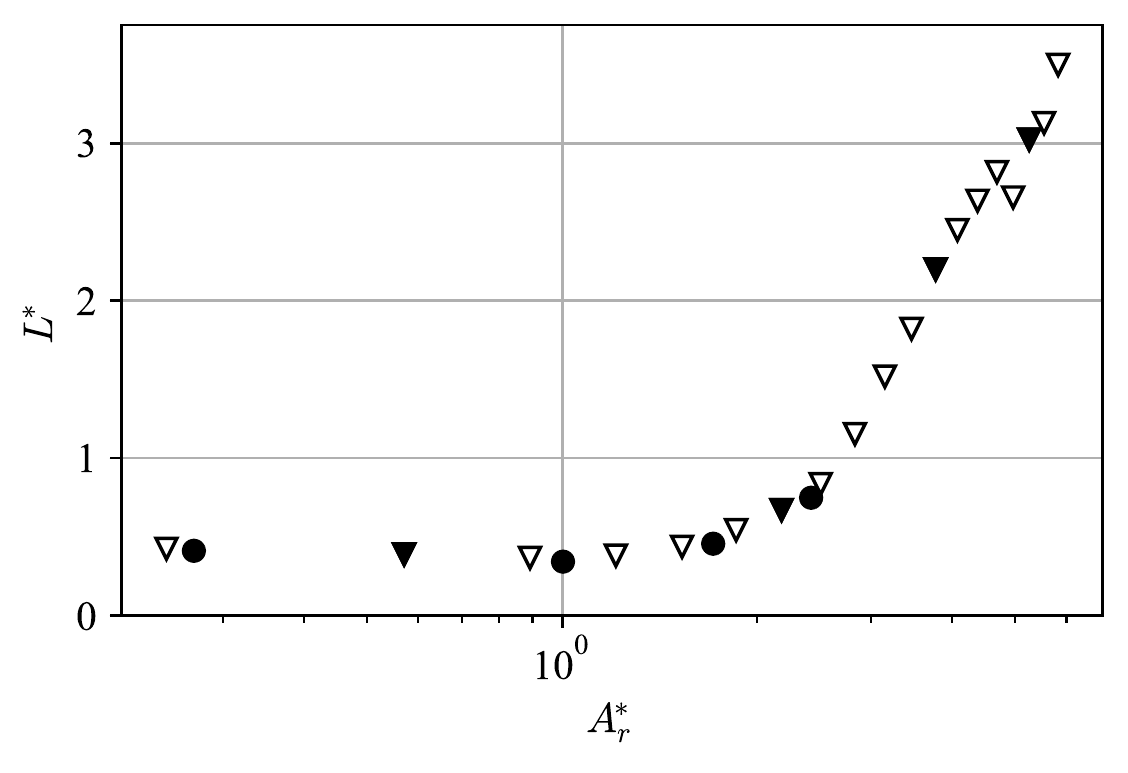}
        \caption{The mean value of the gap $L^*$ as a function of $A_r^*$. The triangular markers have $\delta^*=1/4.5$, $s=7.5$ and $\mu_c=0$. The four solid black circles have $s=2.65$ and otherwise identical simulation settings as the four solid black triangles.}
        \label{fig:sD_Arel_osc_density}
    \end{figure}

    The resulting mean values of the gap are shown in Fig.~\ref{fig:sD_Arel_osc_density}. For a given value of $A^*$, the relative amplitude $A_r^*$ is about a factor two smaller for simulations with a lower density ratio. Even though the value of $A_r^*$ is substantially different at otherwise equal flow conditions, the mean value of the gap falls on top of the data with $s=7.5$. Hence, we find that given the parameters $A_r^*$, $\delta^*$, and $s$, the time-averaged, quasi-steady state of the system (e.g. in terms of $L^*$) is fully determined only by $A_r^*$ and $\delta^*$, (e.g. as in Eq.~\eqref{scalinggap}), and there is no additional dependence on $s$. This supports the conclusion drawn earlier from Eq.~\eqref{particle_power_balance}, that $s$ does not directly influence the mean state of the system. The two dimensionless parameters $A_r^*$ and $\delta^*$ correspond to the two degrees of freedom of the flow, as found in the dimensionless Navier-Stokes equation~\eqref{momentum_dimless}.
    
\begin{figure*}
        \includegraphics{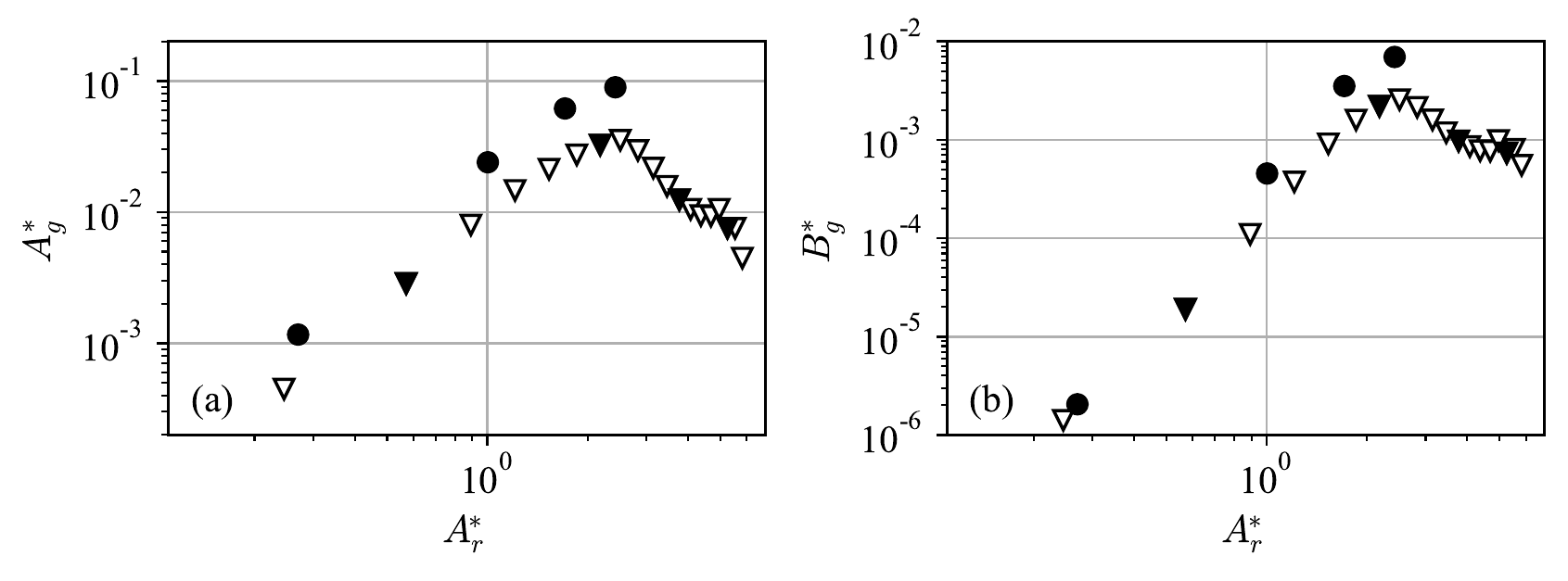}
        \caption{The amplitudes of oscillating gap at two (a) and four (b) times the driving frequency as a function of $A_r^*$. The same coloring is used as in Fig.~\ref{fig:sD_Arel_osc_density}.}
        \label{fig:AgBg_Arel_osc_density}
    \end{figure*}

    In contrast to the collapse for $L^*$, the values of $A_g^*$ and $B_g^*$ for $s=2.56$ and $s=7.5$ do not fall on top of each other as shown in Fig.~\ref{fig:AgBg_Arel_osc_density}. A similar scaling is found, but with proportionality constants approximately a factor two larger. The explanation for this is analogous to the decrease in $A_p^*$ when $s$ is larger. At equal flow conditions the forces exerted on the particles by the steady streaming flow are the same, while the acceleration of the particle is amplified due to the $1/s$ scaling of all terms in Eq.~\eqref{particle_trans_dimless}. So the heavier the particle, the smaller $A_g^*$ and $B_g^*$ are at a given value of $A_r^*$.

\subsection{Effect of particle rotation}\label{results_rotation}
    In the previous sections, we excluded bottom friction to limit the particles' rotation, as observed in experiments by \citeauthor{Klotsa2007}~\cite{Klotsa2007}. In this section, we isolate the effect of the particle rotation on the particle dynamics, the gap size, and steady streaming flows.
    The settings for the simulations with different $\mu_c$-values are $A^*\approx5.4$, $\delta^*=1/3.25$, $s=7.5$, $\Gamma=4.5$ and $\mu_c \in [0.0,1.0]$ with steps of $0.2$.
    
\begin{figure}
        \includegraphics{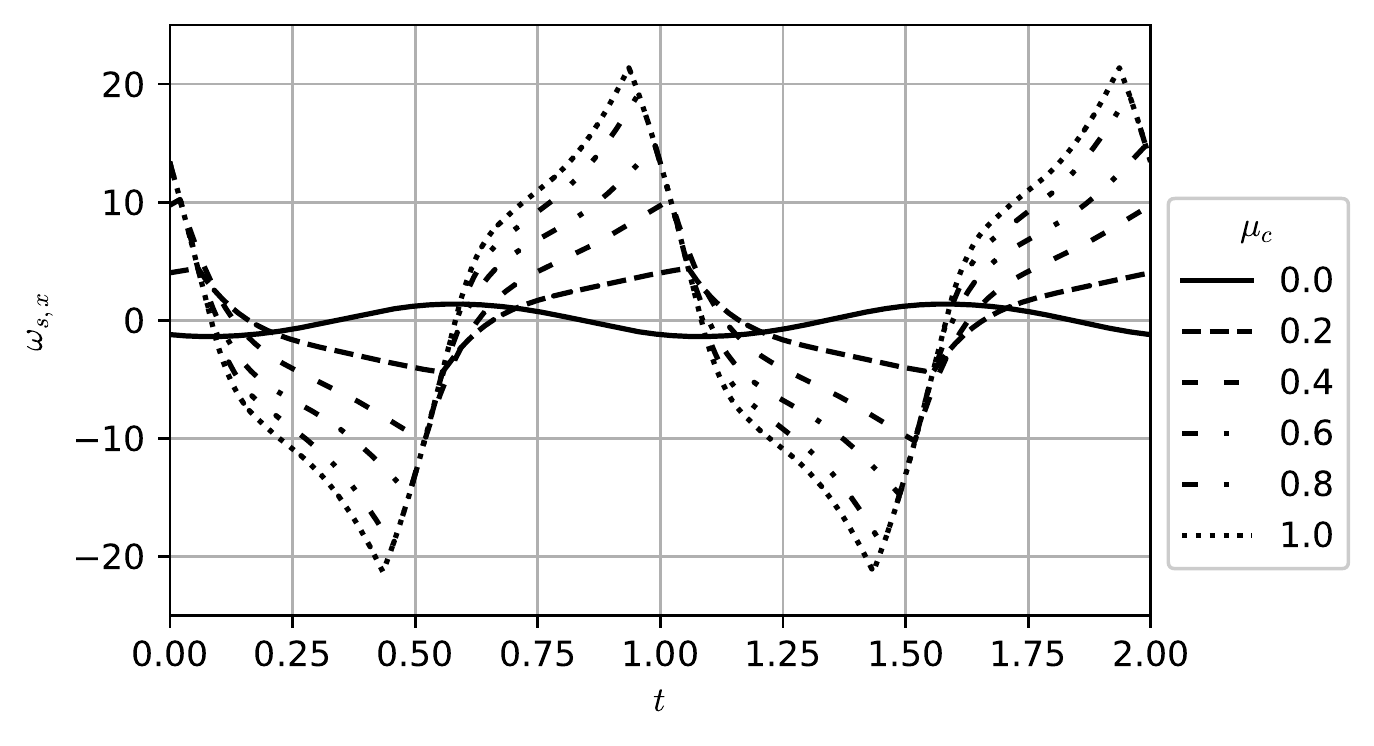}
        \caption{The dimensionless angular velocity of one particle forming a pair as a function of time for simulations with different values of the Coulomb friction coefficient $\mu_c$.}
        \label{fig:wx_muc}
    \end{figure}

    Figure~\ref{fig:wx_muc} shows the effect of varying $\mu_c$ on the particle's (dimensionless) angular velocity. When $\mu_c$ is sufficiently large, the particle rolls `backwards' compared to the case with $\mu_c=0$. This is because the bottom moves underneath the particle, dragging the bottom of the sphere partially with it. The rotation direction is equivalent to the case of a particle rolling over a fixed bottom, which can be seen by moving to the reference frame of the box. The maximum values of the angular velocity $\omega_x$ within one period lie between 1.3 for $\mu_c=0$ and 21.4 for $\mu_c=1.0$.
    
    Despite the major effect on the angular velocity, bottom friction has little influence on $A_r^*$ for the same values of $A^*$ and $\delta^*$. For the particular values of $A^*$, $\delta^*$, $\Gamma$, and $s$, $3.2\alt A_r^* \alt 3.4$ which implies a difference of about $5\%$ between the different $\mu_c$-values. Since the angular velocity is the only quantity changing significantly with changing $\mu_c$ values, it is reasonable to assume that particle rotation is the main cause of changes in $L^*$.
    
\begin{figure}
        \includegraphics{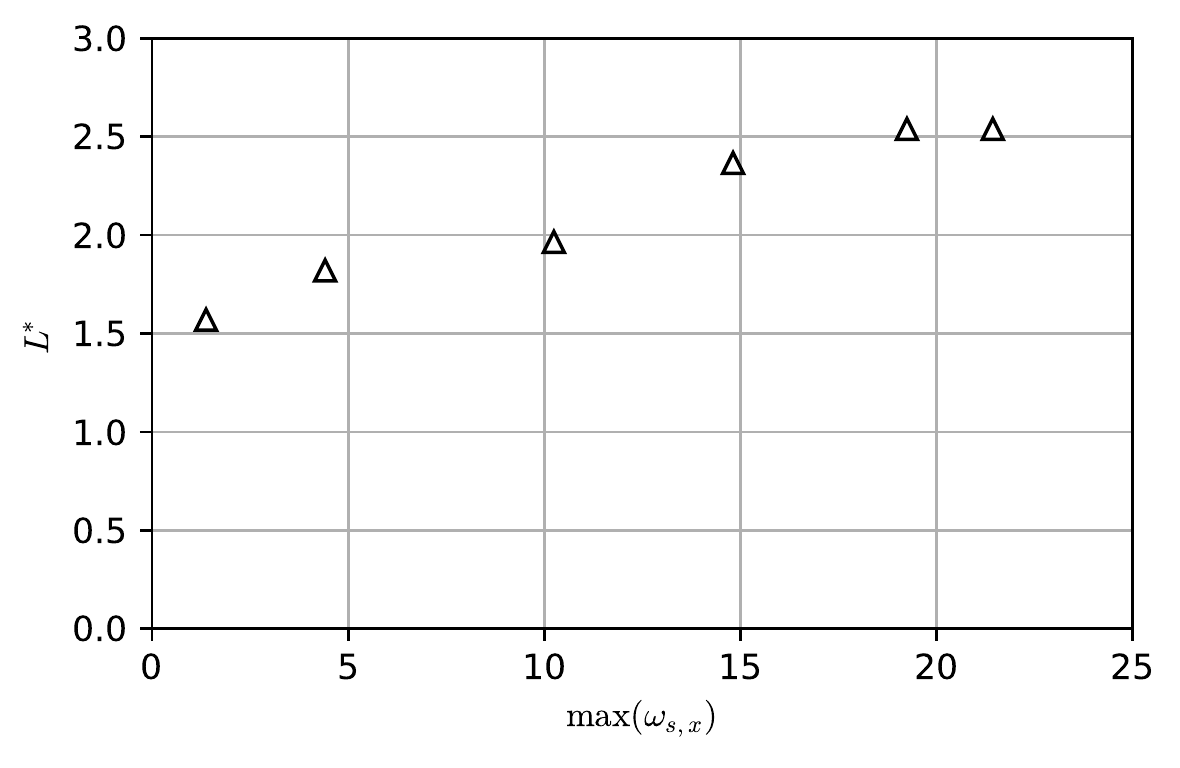}
        \caption{The spacing between particles forming a pair as a function of the maximum angular velocity $\omega_{s,x}$.}
        \label{fig:wxsd}
    \end{figure}

\begin{figure}
        \includegraphics{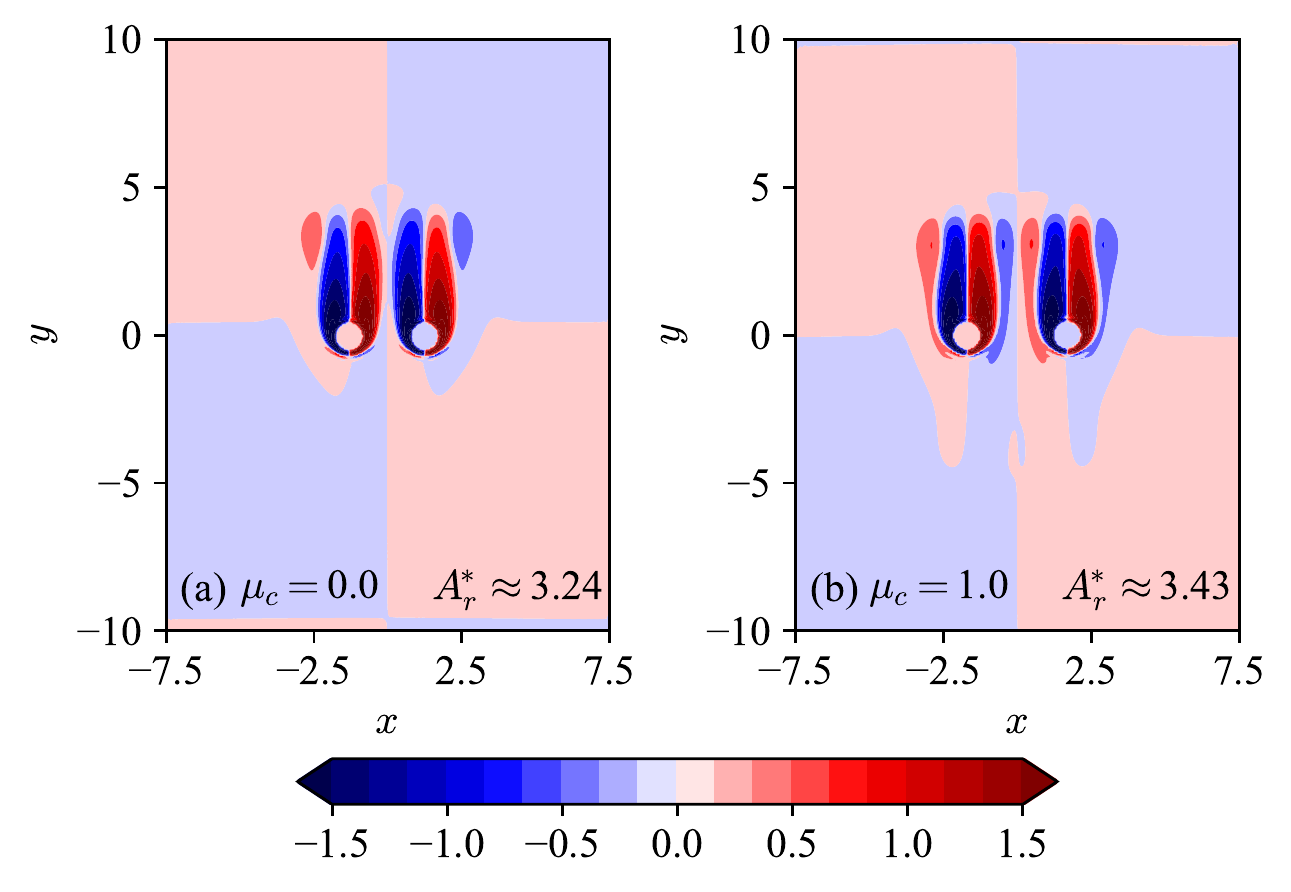}
        \caption{(Color online) The instantaneous vorticity ($\text{sign}(\omega_z)\log\left(\left|\omega_z\right|+1\right)$) at phase $t=0.2$ with $\Gamma=4.5$, $s=7.5$ and $\delta^*=1/3.25$. The Coulomb friction coefficient is either $0$ (a) or $1.0$ (b). The values of $A_r^*$ are given to indicate the small difference in relative amplitudes.}
        \label{fig:VorMu}
    \end{figure}

    Figure~\ref{fig:wxsd} shows the gap size as a function of the maximum dimensionless particle angular velocity. It is clear that particle rotation has a significant effect on the spacing between particles. However, there is little difference in the structure of the averaged vorticity fields, but not in the instantaneous fields, shown in Fig.~\ref{fig:VorMu}. With larger angular rotation, particles are further away from each other and the vorticity distribution around each particle starts to resemble that of an isolated sphere. In Fig.~\ref{fig:VorMu}, an additional dipole-like structure forms between the particles for higher rotation rates. By the time the flow reverses, most of it has been dissipated. This effect is clearest in the video provided as additional material. 
    
\begin{figure}
        \includegraphics{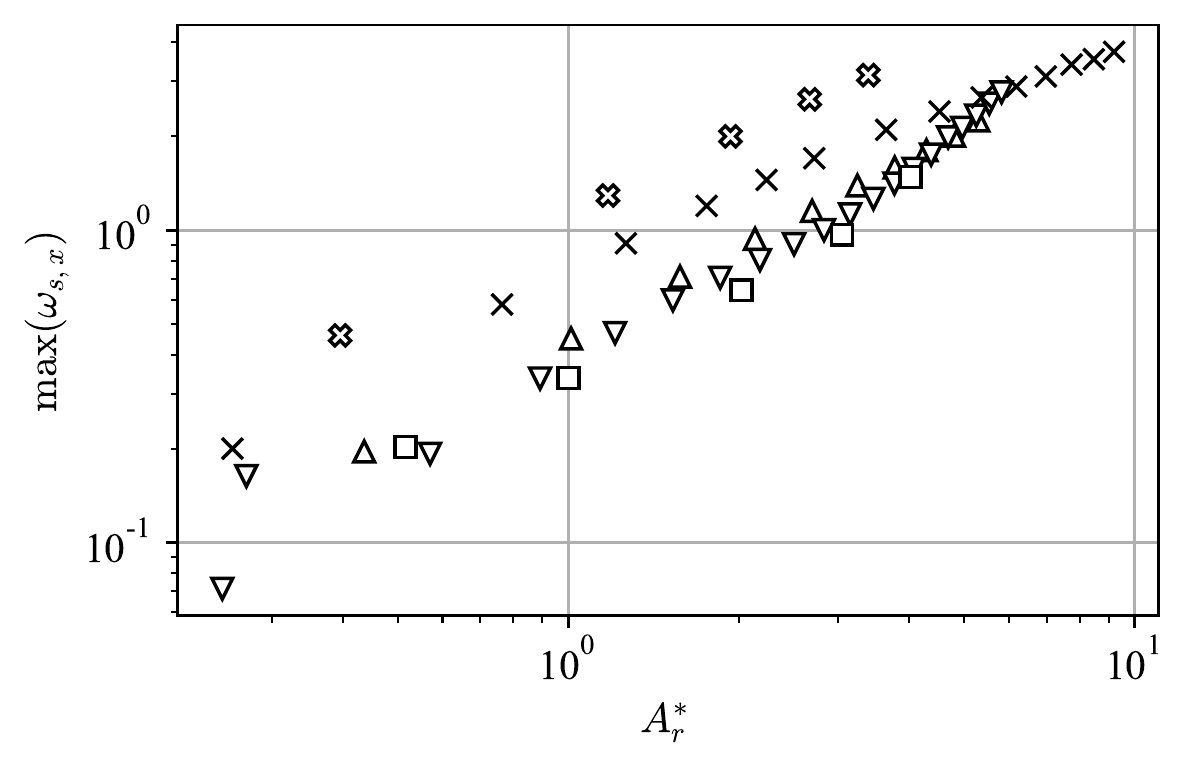}
        \caption{Maximum angular velocities $\omega_{s,x}$ as a function of $A_r^*$ for all simulations with $\mu_c=0$ and $s=7.5$.}
        \label{fig:maxwx}
    \end{figure}

    From Figs.~\ref{fig:wx_muc}\nobreakdash-\ref{fig:wxsd}, we have seen that particle rotation through bottom friction has a significant effect on the value of $L^*$. Even without bottom friction ($\mu_c=0$), there still is some rotation of the particles. To exclude any significant role of rotation on the results from Sec.~\ref{subsec:Quantitative}, we revisit the simulations with $\mu_c=0$ and compare their maximum angular velocities in Fig.~\ref{fig:maxwx}. For the whole data set, $\max(\omega_{s,x})\alt 4$, which is small compared to the range of values in  Fig.~\ref{fig:wxsd}. Furthermore, $\max(\omega_{s,x})\alt 1$ within the viscous regime. It is thus likely that the rotation of the particles does not play an important role in the scaling of the mean gap presented in Eq.~\eqref{scalinggap}, especially not for $A_r^*\leq1$.
   
    All in all, there is an increase in complexity due to the influence of rotation direction and magnitude and the additional parameter $\mu_c/\Gamma$ in the set of dimensionless quantities defining the system. We here refrain from a further in-depth, quantitative analysis on the effect of particle rotation on the spacing of the gap. A good starting point for future studies would be to isolate the effect of rotation on the particle dynamics and steady streaming by examining a system with a single rotating sphere in an oscillating flow, similar to, e.g., \cite{Martin1976}.

\section{\label{sec:Discussion}Discussion}
    A key aspect for the understanding of the numerical results is to relate the particle pair behavior as a function of the different parameters to the governing equations and flow fields.
    For $A_r^*\alt 1$, the pair forms at a distance that depends only on one dimensionless number: $\delta^*$. For these flow conditions, $A^*$ is sufficiently small such that the advection term in Eq.~\eqref{momentum_dimless} is smaller than both the forcing and local acceleration, both of order unity. Flow structures generated by the particle-fluid interactions, such as vortices, are not carried away from the particles but rather stay in their vicinity and are dissipated by viscosity. Clearly, when advection is insignificant, the excursion length of the oscillating flow is irrelevant. However, it is important to note that, even though the advection term is small in the viscous regime, it cannot be neglected. It is the only term that gives a nonzero contribution when the (bulk) flow is averaged over an integer number of periods. Hence, it can be interpreted as the source of the steady streaming flow. Without the non-linearity from the advection, there would be no steady streaming flow to drive the pair formation in the first place.
    
    In an analytical study, \citeauthor{Riley1966}~\cite{Riley1966} calculated the steady streaming flow in the limit of $\delta^*\ll 1$ and $A^*\ll1$. In this case, the vorticity (to first order) is confined to a `shear-wave' region of thickness $O(\delta^*)$. If this region sets the minimum distance between the two particles, it provides a suitable explanation for our viscous regime. 
    The theory by \citeauthor{Riley1966} also means that $L^*\rightarrow 0$ in the limit $\delta^*\rightarrow0$, which matches with Fig.~\ref{fig:sD_Arel_osc} and the scaling of Eq.~\eqref{scalinggap}. The existence of a minimum gap value is further supported by recent numerical findings in the limit of $A_r\rightarrow0$ \cite{Jalal2016,Fabre2017}.

    Despite the fact that increasing $A_r^*$ has no effect on the mean value of the gap in the viscous regime, the amplitude of the oscillation of the gap, through $A_g^*$ and $B_g^*$, does vary strongly with the quadratic and quartic scalings of Eq.~\eqref{AB_osc_scaling}, respectively. This can be explained by the fact that while the advection of vortices may be small, their strength does increase with $A_r^*$. The vorticity is generated by the shear flow near the no-slip surfaces of the particles, which in turn increases with larger relative velocities ($\sim A_r\omega$). The change in vorticity magnitude is clearly visible in the first few plots of Fig.~\ref{fig:steadystreamingvort}. The stronger vortices that stay in vicinity of the pair cause larger forces on the particles. Hence, in the viscous regime a change in $A_r^*$ results in stronger attractive and repulsive particle pair interactions and thus higher oscillation amplitudes $A_g^*$ and $B_g^*$. 
    
    Conversely, the lower oscillation amplitudes in the advective regime, as illustrated by the negative slope of $A_g^*$ in Fig.~\ref{fig:Agap_Arel_osc}, can be linked to the widening of the gap. The gap grows faster than the vorticity magnitude, and thus, the strength of particle-fluid interactions is diminished.

    Around the transition between the regimes, $A_r^*$ is of order unity, and the advection term becomes increasingly important compared to the forcing in Eq.~\eqref{momentum_dimless}. The vorticity that is initially concentrated around the particles gets advected before the flow direction reverses. It is likely that this is enhanced by the formation of the dipole-like vorticity structures between the pair that propagate with respect to the flow. The larger the value of $A_r$, the higher the dipole strength and propagation velocity are. The dipole may then travel away from the particles before it dissipates, similarly to the propagation and escape of a dipole away from a tidal inlet \cite{Wells2003}. This additional effect could explain the increasingly long vorticity patches between the pair, as observed in Fig.~\ref{fig:steadystreamingvort}. It also matches the observation by \citeauthor{Klotsa2007}~\cite{Klotsa2007} of a stagnation point that moves away from the pair at higher amplitudes.
    
    The gap between the particles widens in the advective regime. In this regime, instead of forming a strong entity, the particles seem weakly coupled to each other by the relatively weak outer part of the steady streaming flow. This goes hand in hand with the steady streaming flow resembling that around two isolated particles. However, the causality between the gap widening and flow changes is not straightforward to describe due to the non-linearity in the two-way coupling between the flow and the particles. The topological changes in the flow and the increase of $L^*$ come together and one cannot be attributed to the other. We can only conclude that the increase in the non-linear interactions in the flow between the particles results in an increase of $L^*$.
    
    Moreover, the large gap and weak particle-particle interactions in the advective regime cause a greater uncertainty in the quantities describing the gap. As the instantaneous forces between the particles decrease, other smaller effects, such as self-interactions through the periodic boundaries or under-resolved flow structures, can cause a significant change in e.g. the mean gap.
    
    The improved agreement with experimental data with respect to the comparison in the original paper \cite{Klotsa2007} is most probably due to a more physically realistic approach in our numerical method. In the original study, the particles moved in the $y$-direction along a predetermined path in a fluid that is at rest. However, the simulations presented in the current paper show a more complex particle motion, with the trajectories describing an elongated \emph{figure 8}. Additionally, our numerical method incorporates lubrication forces between the particles and bottom. In fact, \citeauthor{Klotsa2007}~\cite{Klotsa2007} already identified the small region between the particle and the bottom of the domain as a part in their numerical approach that could be improved. 

\section{\label{sec:Conclusions}Conclusions}
    In this paper, we describe the dynamics of submerged particles forming pairs inside an oscillating box. The results are obtained using direct numerical simulations where the Navier-Stokes equations are solved in a double-periodic domain with the top and bottom oscillating with the flow. The particles are simulated using an immersed boundary method. Our simulations clearly show the formation of stable particle pairs with the particles oscillating both parallel and perpendicular to the oscillating flow. In other words, they describe elongated \emph{figure 8} trajectories that, to the best of our knowledge, have not been previously reported.
    
    The system is described by four non-dimensional parameters: the normalized excursion length of the box $A^*$, the normalized Stokes boundary layer thickness $\delta^*$, the density ratio $s$, and the ratio $\mu_c/\Gamma$ where $\mu_c$ is the Coulomb friction coefficient and $\Gamma$ is the non-dimensional acceleration. Simulations with $\mu_c=0$ show that the size of the gap between the spheres depends on only two parameters: $\delta^*$ and $A_r^*$. The latter is the normalized relative excursion length of the particles with respect to the fluid and is equivalent to the Keulegan-Carpenter number. In addition, it is proportional to $A^*$ with the proportionality constant depending on $\delta^*$, $s$ and $\mu_c$. As long as $\mu_c=0$, variation of $s$ affects only this proportionality constant.
    
    Both the mean gap and its (spanwise) oscillations show two regimes: a viscous and an advective regime. At low amplitudes ($A_r^*\alt 1$), the mean particle separation only depends on $\delta^*$ and the magnitude of the oscillations increases with $A_r^*$. At higher amplitudes ($A_r^*\agt 2$), the gap widens with ${A_r^*}^3$ and the magnitude of the oscillations decreases. The experimental data from \citeauthor{Klotsa2007}~\cite{Klotsa2007} further supports the results for the mean gap. Additionally, we found that the Reynolds number based on the mean gap $\mathrm{Re}_L$ depends only on the single parameter $A_r^*/{\delta^*}^{0.5}$, while still exhibiting the two regimes and transition between them.
    
    The effect of particle rotation is isolated by varying the bottom friction coefficient. The angular velocity within each period varies complexly in time. Based on only the maximum angular velocity, we showed that with more rotation, the gap widens and the flow fields around the particle pairs get more complex structures.
    
    Our results suggest that the motion of the particles perpendicular to the main oscillation, i.e. the oscillation of the gap, has a dynamical importance. In quasi-steady state, the forces on the particles may be zero on average, but they oscillate significantly within each period. An accurate description of the particle-fluid interaction within each period and on the sub-particle level is here of paramount importance. The results from numerical studies in which the particles are confined in one \cite{Klotsa2007} or all \cite{Jalal2016} directions can thus not explain the particle behavior in systems where they are free to move. The good agreement with the experimental data validates that the used method is suitable for more extensive future studies. 
    The new and revised insight of the particle pairs provided in this paper is an important step towards the understanding of denser systems such as particle chains and more complex systems such as model swimmers. 
    
\begin{acknowledgments}
    We like to thank the staff in charge of the Reynolds cluster at the Delft University of Technology. 
\end{acknowledgments}

\appendix*

\section{\label{sec:Fitting}Equations used for least square fittings}
    The relative particle amplitude $A_r^*$ is obtained by applying a least-squares fit to the particle position in the $y$\nobreakdash-direction. The used function is given by
    \begin{equation}
        y(t) = y_0 + v_0t + A_r^*\sin{(2\pi t+\theta)},
    \end{equation}
    where $y_0$ and $v_0$ are fitting constants. These constants correct for remaining transient effects such as offsets or drifts in the particle trajectory.
    
    The mean amplitude of the gap $L^*$ is obtained from the distance between the particles in the $x$\nobreakdash-direction, using either the fitting function
    \begin{eqnarray}
        x(t) = &L^* + a e^{-t/\tau} + A\cos(4\pi t+\theta_1) \nonumber\\
        &+ B\cos(8\pi t+\theta_2) + C\cos(2\pi t+\theta_3)
    \end{eqnarray}
    or 
    \begin{eqnarray}
       x(t) = &L^* - \frac{1}{b+ct} + A\cos(4\pi t+\theta_1) \nonumber\\
       &+ B\cos(8\pi t+\theta_2) + C\cos(2\pi t+\theta_3).
    \end{eqnarray}
    Based on the quality of the least-squares fit, either the inverse proportionality or exponentially decaying function is used.

    The normalized amplitudes of the oscillation of the gap $A_g^*$, $B_g^*$ and $C_g^*$ are obtained in a similar manner as $L^*$, now using the fitting function
    \begin{eqnarray}
        x(t) = &a+bt+ct^2 + A_g^*\cos(4\pi t+\theta_1) \nonumber\\
        &+ B_g^*\cos(8\pi t+\theta_2) + C_g^*\cos(2\pi t+\theta_3),
    \end{eqnarray}
    where we do not assign any meaning to the polynomial part, as it is only used for interpolation over a few periods of the main oscillation. The three harmonics are interpret as follows. The gap oscillates with amplitude $A_g^*$ and twice the frequency of the oscillating flow. So back and forth have identical contributions. The magnitude of the asymmetry in this oscillation is given by $B_g^*$. 
    Changes in the oscillation with the same frequency as the flow oscillation scale with $C_g^*$.


%

\end{document}